\documentclass[12pt,a4paper]{article}
\pdfoutput=1

\usepackage{ifthen} 
\newboolean{pdflatex}
\setboolean{pdflatex}{true} 

\newboolean{articletitles}
\setboolean{articletitles}{true} 

\newboolean{uprightparticles}
\setboolean{uprightparticles}{false} 

\def\paperauthors{LHCb collaboration} 
\def\paperasciititle{Measurement of the electron reconstruction efficiency at LHCb}
\def\papertitle{Measurement of the electron reconstruction efficiency at LHCb}
\def\paperkeywords{{High Energy Physics}, {LHCb}}
\def\papercopyright{\the\year\ CERN for the benefit of the LHCb collaboration} 
\def\paperlicence{CC-BY-4.0 licence}
\def\paperlicenceurl{https://creativecommons.org/licenses/by/4.0/}

\usepackage[top=1in, bottom=1.25in, left=1in, right=1in]{geometry}

\usepackage{microtype}
\usepackage{lineno}  
\usepackage{xspace} 
\usepackage{caption} 

\usepackage{graphicx}  
\usepackage{color}
\usepackage{colortbl}
\graphicspath{{./figs/}} 
\DeclareGraphicsExtensions{.pdf,.PDF,png,.PNG}

\usepackage{amsmath} 
\usepackage{amssymb}
\usepackage{amsfonts}
\usepackage{upgreek} 

\newcommand*\patchAmsMathEnvironmentForLineno[1]{%
\expandafter\let\csname old#1\expandafter\endcsname\csname #1\endcsname
\expandafter\let\csname oldend#1\expandafter\endcsname\csname
end#1\endcsname
 \renewenvironment{#1}%
   {\linenomath\csname old#1\endcsname}%
   {\csname oldend#1\endcsname\endlinenomath}%
}
\newcommand*\patchBothAmsMathEnvironmentsForLineno[1]{%
  \patchAmsMathEnvironmentForLineno{#1}%
  \patchAmsMathEnvironmentForLineno{#1*}%
}
\AtBeginDocument{%
\patchBothAmsMathEnvironmentsForLineno{equation}%
\patchBothAmsMathEnvironmentsForLineno{align}%
\patchBothAmsMathEnvironmentsForLineno{flalign}%
\patchBothAmsMathEnvironmentsForLineno{alignat}%
\patchBothAmsMathEnvironmentsForLineno{gather}%
\patchBothAmsMathEnvironmentsForLineno{multline}%
\patchBothAmsMathEnvironmentsForLineno{eqnarray}%
}

\usepackage{hyperxmp}

\usepackage[pdftex,
            pdfauthor={\paperauthors},
            pdftitle={\paperasciititle},
            pdfkeywords={\paperkeywords},
            pdfcopyright={Copyright (C) \papercopyright},
            pdflicenseurl={\paperlicenceurl}]{hyperref}

\usepackage[colorinlistoftodos,textsize=scriptsize]{todonotes}

\usepackage[all]{hypcap} 

\usepackage{xspace} 
\usepackage{upgreek}

\def\lhcb   {\mbox{LHCb}\xspace}

\def\velo   {VELO\xspace}

\def\MagUp {\mbox{\em Mag\kern -0.05em Up}\xspace}

\ifthenelse{\boolean{uprightparticles}}%
{

 \def\Pmu         {\ensuremath{\upmu}\xspace}                 
 \def\Pnu         {\ensuremath{\upnu}\xspace}

 \def\Ppsi        {\ensuremath{\uppsi}\xspace}

 \def\PDelta      {\ensuremath{\Delta}\xspace}                 
 \def\PXi         {\ensuremath{\Xi}\xspace}                 
 \def\PLambda     {\ensuremath{\Lambda}\xspace}                 
 \def\PSigma      {\ensuremath{\Sigma}\xspace}                 
 \def\POmega      {\ensuremath{\Omega}\xspace}                 
 \def\PUpsilon    {\ensuremath{\Upsilon}\xspace}

 \def\PB      {\ensuremath{\mathrm{B}}\xspace}                 
                  
 \def\PD      {\ensuremath{\mathrm{D}}\xspace}

 \def\PJ      {\ensuremath{\mathrm{J}}\xspace}                 
 \def\PK      {\ensuremath{\mathrm{K}}\xspace}

 \def\Pb      {\ensuremath{\mathrm{b}}\xspace}                 
 \def\Pc      {\ensuremath{\mathrm{c}}\xspace}                 
                  
 \def\Pe      {\ensuremath{\mathrm{e}}\xspace}

 \def\Pi      {\ensuremath{\mathrm{i}}\xspace}

 \def\thebaroffset{0.0em}
}
{

 \def\Pmu         {\ensuremath{\mu}\xspace}                 
 \def\Pnu         {\ensuremath{\nu}\xspace}

 \def\Ppsi        {\ensuremath{\psi}\xspace}                 
                  
 \mathchardef\PDelta="7101
 \mathchardef\PXi="7104
 \mathchardef\PLambda="7103
 \mathchardef\PSigma="7106
 \mathchardef\POmega="710A
 \mathchardef\PUpsilon="7107
                  
 \def\PB      {\ensuremath{B}\xspace}                 
                  
 \def\PD      {\ensuremath{D}\xspace}

 \def\PJ      {\ensuremath{J}\xspace}                 
 \def\PK      {\ensuremath{K}\xspace}

 \def\Pb      {\ensuremath{b}\xspace}                 
 \def\Pc      {\ensuremath{c}\xspace}                 
                  
 \def\Pe      {\ensuremath{e}\xspace}

 \def\Pi      {\ensuremath{i}\xspace}

 \def\thebaroffset{0.18em}
}
\newcommand{\offsetoverline}[2][\thebaroffset]{\kern #1\overline{\kern -#1 #2}}%

\makeatletter
\ifcase \@ptsize \relax
  \newcommand{\miniscule}{\@setfontsize\miniscule{4}{5}}
\or
  \newcommand{\miniscule}{\@setfontsize\miniscule{5}{6}}
\or
  \newcommand{\miniscule}{\@setfontsize\miniscule{5}{6}}
\fi
\makeatother

\DeclareRobustCommand{\optbar}[1]{\shortstack{{\miniscule (\rule[.5ex]{1.25em}{.18mm})}
  \\ [-.7ex] $#1$}}

\def\en         {{\ensuremath{\Pe^-}}\xspace}   
\def\ep         {{\ensuremath{\Pe^+}}\xspace}

\def\mup        {{\ensuremath{\Pmu^+}}\xspace}
\def\mun        {{\ensuremath{\Pmu^-}}\xspace} 

\def\ellell     {\ensuremath{\ell^+ \ell^-}\xspace}

\def\neub       {{\ensuremath{\overline{\Pnu}}}\xspace}

\def\neulb      {{\ensuremath{\neub_\ell}}\xspace}

\def\cquark    {{\ensuremath{\Pc}}\xspace}

\def\bquark    {{\ensuremath{\Pb}}\xspace}

\def\kaon    {{\ensuremath{\PK}}\xspace}

\def\KorKbar {\kern \thebaroffset\optbar{\kern -\thebaroffset \PK}{}\xspace}

\def\Kp      {{\ensuremath{\kaon^+}}\xspace}

\def\Kstar   {{\ensuremath{\kaon^*}}\xspace}

\def\D       {{\ensuremath{\PD}}\xspace}

\def\DorDbar {\kern \thebaroffset\optbar{\kern -\thebaroffset \PD}\xspace}

\def\Dp      {{\ensuremath{\D^+}}\xspace}

\def\B       {{\ensuremath{\PB}}\xspace}

\def\BorBbar {\kern \thebaroffset\optbar{\kern -\thebaroffset \PB}\xspace}
\def\Bz      {{\ensuremath{\B^0}}\xspace}

\def\Bu      {{\ensuremath{\B^+}}\xspace}

\def\Bp      {{\ensuremath{\Bu}}\xspace}

\def\jpsi     {{\ensuremath{{\PJ\mskip -3mu/\mskip -2mu\Ppsi\mskip 2mu}}}\xspace}

\def\Y#1S{\ensuremath{\PUpsilon{(#1S)}}\xspace}

\def\LorLbar     {\kern \thebaroffset\optbar{\kern -\thebaroffset \PLambda}\xspace}

\def\to                 {\ensuremath{\rightarrow}\xspace}

\def\order   {{\ensuremath{\mathcal{O}}}\xspace}

\def\AT#1     {\ensuremath{A_{\mathrm{T}}^{#1}}\xspace}           

\def\C#1      {\ensuremath{\mathcal{C}_{#1}}\xspace}                       
\def\Cp#1     {\ensuremath{\mathcal{C}_{#1}^{'}}\xspace}                    
\def\Ceff#1   {\ensuremath{\mathcal{C}_{#1}^{\mathrm{(eff)}}}\xspace}        
\def\Cpeff#1  {\ensuremath{\mathcal{C}_{#1}^{'\mathrm{(eff)}}}\xspace}       
\def\Ope#1    {\ensuremath{\mathcal{O}_{#1}}\xspace}                       
\def\Opep#1   {\ensuremath{\mathcal{O}_{#1}^{'}}\xspace}                    

\newcommand{\nospaceunit}[1]{\ensuremath{\text{#1}}}       
\newcommand{\aunit}[1]{\ensuremath{\text{\,#1}}}       

\newcommand{\tev}{\aunit{Te\kern -0.1em V}\xspace}
\newcommand{\gev}{\aunit{Ge\kern -0.1em V}\xspace}
\newcommand{\mev}{\aunit{Me\kern -0.1em V}\xspace}
\newcommand{\kev}{\aunit{ke\kern -0.1em V}\xspace}
\newcommand{\ev}{\aunit{e\kern -0.1em V}\xspace}
\newcommand{\mevc}{\ensuremath{\aunit{Me\kern -0.1em V\!/}c}\xspace}
\newcommand{\gevc}{\ensuremath{\aunit{Ge\kern -0.1em V\!/}c}\xspace}
\newcommand{\mevcc}{\ensuremath{\aunit{Me\kern -0.1em V\!/}c^2}\xspace}
\newcommand{\gevcc}{\ensuremath{\aunit{Ge\kern -0.1em V\!/}c^2}\xspace}

\def\mm   {\aunit{mm}\xspace}

\def\mum  {\ensuremath{\,\upmu\nospaceunit{m}}\xspace}

\def\fb   {\ensuremath{\aunit{fb}}\xspace}
\def\invfb   {\ensuremath{\fb^{-1}}\xspace}

\def\order{{\ensuremath{\mathcal{O}}}\xspace}

\def\gsim{{~\raise.15em\hbox{$>$}\kern-.85em
          \lower.35em\hbox{$\sim$}~}\xspace}
\def\lsim{{~\raise.15em\hbox{$<$}\kern-.85em
          \lower.35em\hbox{$\sim$}~}\xspace}

\def\pt         {\ensuremath{p_{\mathrm{T}}}\xspace}

\def\ptot       {\ensuremath{p}\xspace}

\def\mrad{\aunit{mrad}}

\def\evtgen     {\mbox{\textsc{EvtGen}}\xspace}

\def\geant      {\mbox{\textsc{Geant4}}\xspace}

\def\photos     {\mbox{\textsc{Photos}}\xspace}

\def\pythia     {\mbox{\textsc{Pythia}}\xspace}

\xspace

\def\tell1  {TELL1\xspace}
\def\ukl1   {UKL1\xspace}

\newcommand{\eg}{\mbox{\itshape e.g.}\xspace}
\newcommand{\ie}{\mbox{\itshape i.e.}\xspace}

\usepackage{cite} 
\usepackage{mciteplus}
\usepackage[all]{nowidow}

\begin{document}

\renewcommand{\thefootnote}{\fnsymbol{footnote}}
\setcounter{footnote}{1}


\begin{titlepage}
\pagenumbering{roman}

\vspace*{-1.5cm}
\centerline{\large EUROPEAN ORGANIZATION FOR NUCLEAR RESEARCH (CERN)}
\vspace*{1.5cm}
\noindent
\begin{tabular*}{\linewidth}{lc@{\extracolsep{\fill}}r@{\extracolsep{0pt}}}
\ifthenelse{\boolean{pdflatex}}
{\vspace*{-1.5cm}\mbox{\!\!\!\includegraphics[width=.14\textwidth]{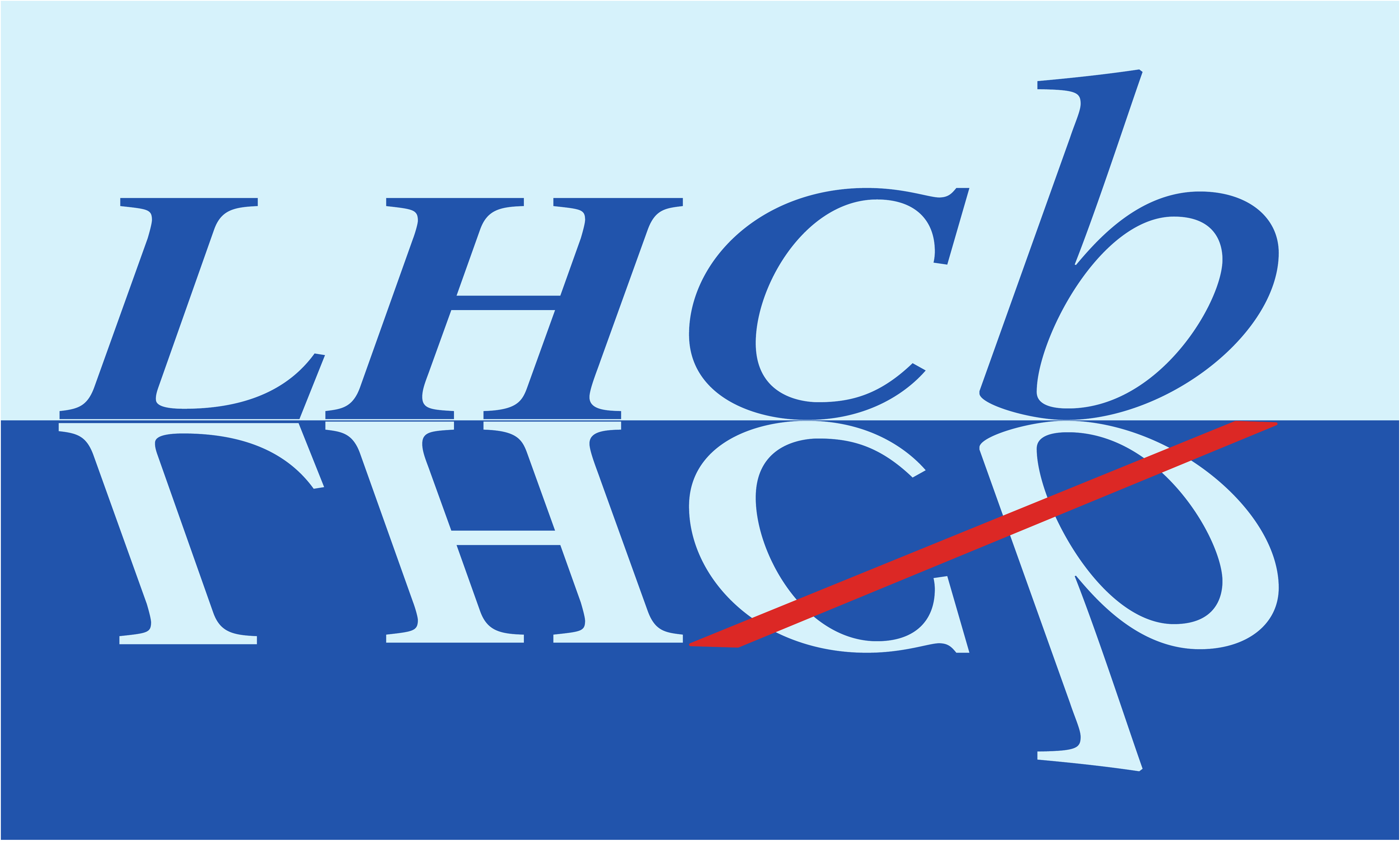}} & &}%
{\vspace*{-1.2cm}\mbox{\!\!\!\includegraphics[width=.12\textwidth]{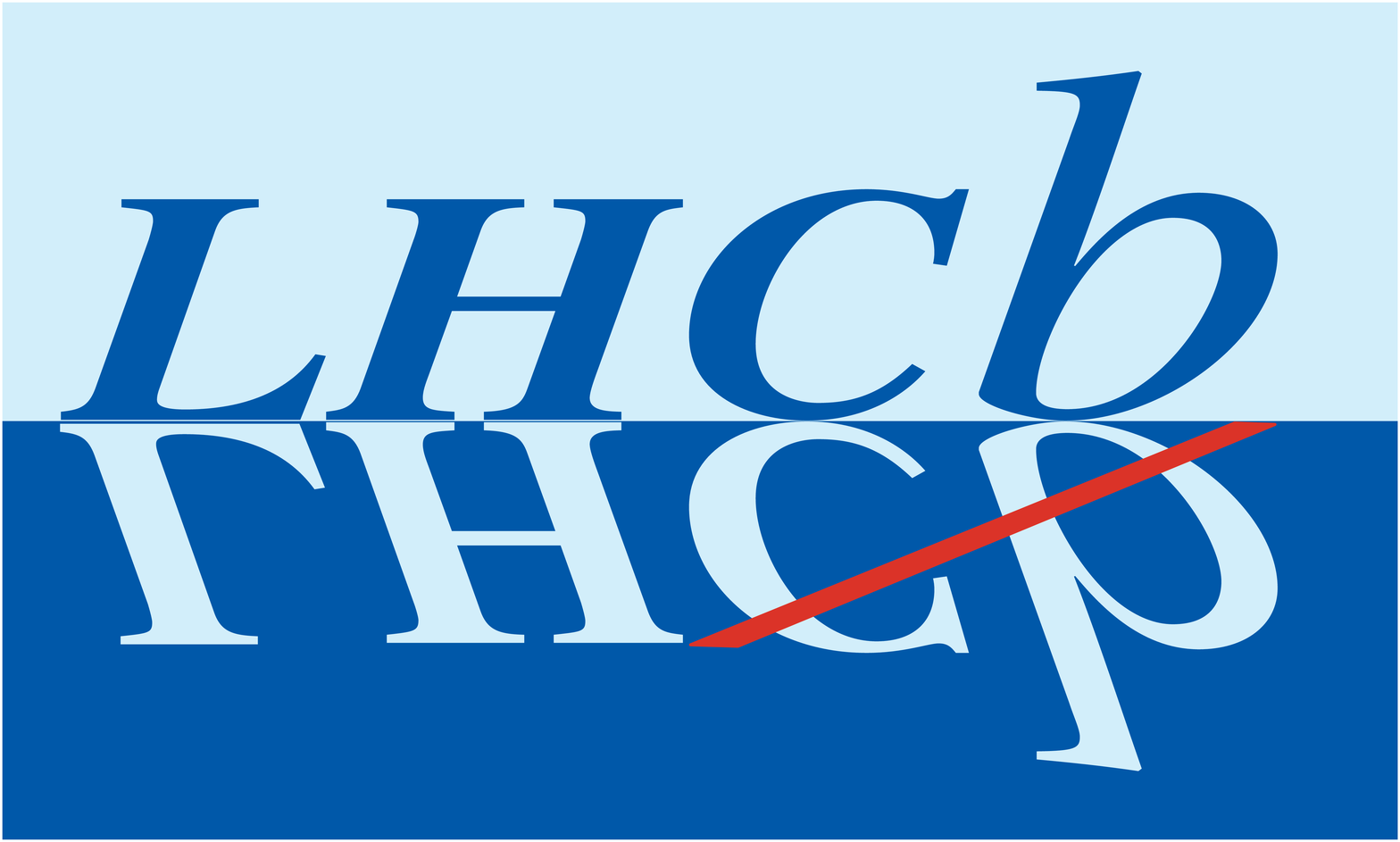}} & &}%
\\
 & & CERN-EP-2019-181 \\  
 & & LHCb-DP-2019-003 \\  
 & & November 25, 2019 \\ 
 & & \\
\end{tabular*}

\vspace*{4.0cm}

{\normalfont\bfseries\boldmath\huge
\begin{center}
  \papertitle 
\end{center}
}

\vspace*{2.0cm}

\begin{center}
\paperauthors\footnote{Authors are listed at the end of this paper.}
\end{center}

\vspace{\fill}

\begin{abstract}
  \noindent
    The single electron track-reconstruction efficiency is calibrated using a sample corresponding to $1.3$~\invfb of $pp$ collision data recorded with the LHCb detector in 2017.
  This measurement exploits $\Bp\to\jpsi(\ep\en)\Kp$ decays, where one of the electrons
  is fully reconstructed and paired with the kaon, while the other electron is reconstructed using only the information of the vertex detector. Despite this partial reconstruction, kinematic and geometric constraints allow the \B meson mass to be reconstructed and the signal to be well separated from backgrounds. This in turn allows the electron reconstruction efficiency to be measured by matching the partial track segment found in the vertex detector to tracks found by LHCb's regular reconstruction algorithms. The agreement between data and simulation is evaluated, and corrections are derived for simulated electrons in bins of kinematics. These correction factors allow LHCb to measure branching fractions involving single electrons with a systematic uncertainty below $1\%$.
  
\end{abstract}

\vspace*{2.0cm}

\begin{center}
  Published in JINST 14 (2019) P11023
\end{center}

\vspace{\fill}

{\footnotesize 

\centerline{\copyright~\papercopyright. \href{\paperlicenceurl}{\paperlicence}.}}
\vspace*{2mm}

\end{titlepage}


\newpage
\setcounter{page}{2}
\mbox{~}

\cleardoublepage

\renewcommand{\thefootnote}{\arabic{footnote}}
\setcounter{footnote}{0}

\pagestyle{plain} 
\setcounter{page}{1}
\pagenumbering{arabic}

\section{Introduction}
\label{sec:Introduction}
The \lhcb detector is designed for the study of \bquark-hadron and \cquark-hadron decays, using proton-proton collisions provided by the LHC. In order to achieve these physics goals, the trajectories of stable charged particles (tracks) need to be reconstructed accurately and with a high and well-known efficiency. The precise knowledge of this charged-particle reconstruction efficiency is essential in measurements of branching fractions, for example. 

While muons produced in heavy flavour decays are nearly unhindered by the \lhcb detector material, electrons suffer from energy loss via bremsstrahlung. In some cases, the energy loss affects the trajectory to such a degree that the electrons no longer traverse all tracking detectors. Therefore, the track-reconstruction efficiency for electrons describes not only the performance of the track-reconstruction algorithms for electrons, but also the impact of this loss in geometrical acceptance. 

Thus far, no direct measurement of the electron reconstruction efficiency at \lhcb exists. Tests of lepton universality in $b\to s \ell\ell$ ($\ell=\mu, e$) transitions (\eg Ref.~\cite{LHCb-PAPER-2014-024,LHCb-PAPER-2017-013,LHCb-PAPER-2019-009}) have estimated the efficiency difference between muons and electrons using simulation. This estimate is validated by measuring the ratio of branching fractions $\mathcal{B}\left(\jpsi\to e^+ e^-\right)/\mathcal{B}\left(\jpsi \to\mup \mun\right)$, which is constrained by theory and external measurements to unity. A measurement of the single electron efficiency is not only a complementary check to this method, but also enables tests of lepton universality in decays producing a single electron, such as $\Bz\to\Dp \ell^- \neulb$ decays.

This paper presents a new method to LHCb which combines a tag-and-probe approach with kinematic constraints to determine the electron reconstruction efficiency. Part of the efficiency, which includes the majority of the acceptance losses, is extracted from data. This is achieved by using $\Bp\to\jpsi(\ep\en)\Kp$ decays\footnote{Charge-conjugate modes are implied throughout this paper.}, where one of the electrons and the kaon (the tag) are reconstructed using all standard tracking detectors, while for the other electron (the probe) only the information of the vertex detector is used in its reconstruction. Finally, simulation is used to account for the reconstruction efficiency in the vertex detector. The accuracy of the measured reconstruction efficiency is confirmed in simulation by comparing its results to generator-level quantities, and systematic uncertainties are assigned for residual inaccuracies. 
\section{Detector and simulation}
\label{sec:Detector}
The \lhcb detector~\cite{LHCb-DP-2008-001,LHCb-DP-2014-002} is a single-arm forward
spectrometer covering the \mbox{pseudorapidity} range $2<\eta <5$. 
The detector includes a high-precision tracking system
consisting of a silicon-strip vertex detector, called the \velo, surrounding the $pp$
interaction region~\cite{LHCb-DP-2014-001}, a large-area silicon-strip detector, the TT, located
upstream of a dipole magnet with a bending power of about
$4{\mathrm{\,Tm}}$; and three stations of silicon-strip detectors (Inner Tracker) and straw
drift tubes (Outer Tracker)~\cite{LHCb-DP-2017-001} placed downstream of the magnet, referred to as the T stations.
The complete tracking system provides a measurement of the momentum, \ptot, of charged particles with
a relative uncertainty that varies from 0.5\% at low momentum to 1.0\% at 200\gevc.
The minimum distance of a track to a primary vertex (PV), the impact parameter (IP), 
is measured with a resolution of $(15+29/\pt)\mum$,
where \pt is the component of the momentum transverse to the beam, in\,\gevc. The IP is exploited to discriminate prompt tracks from heavy flavour decays. The dipole magnet deflects oppositely charged particles in opposite
directions. Its polarity is periodically
reversed throughout the data-taking. The configuration with the magnetic field pointing upwards (downwards), 
 bends positively (negatively) charged particles
in the horizontal plane towards the centre of the LHC ring. 
Different types of charged hadrons are distinguished using information
from two ring-imaging Cherenkov detectors~\cite{LHCb-DP-2012-003}. 
Photons, electrons and hadrons are identified by a calorimeter system consisting of
scintillating-pad and preshower detectors, an electromagnetic
calorimeter and a hadronic calorimeter. Muons are identified by a
system composed of alternating layers of iron and multiwire
proportional chambers~\cite{LHCb-DP-2012-002}. 

During Run 2 of the LHC, LHCb operated at $\mu=1.1$, where $\mu$ is defined as the average number of visible interactions per beam-beam crossing. The resulting primary vertices are reconstructed using a seeding and fitting procedure, optimised for vertices distributed nearby the beam line~\cite{Kucharczyk:1756296}. The resolution of the reconstructed PVs depends on the track multiplicity of the PV is presented in Ref.~\cite{LHCb-DP-2019-001}. 

\begin{figure}[]\centering
	   \includegraphics[width=0.78\textwidth]{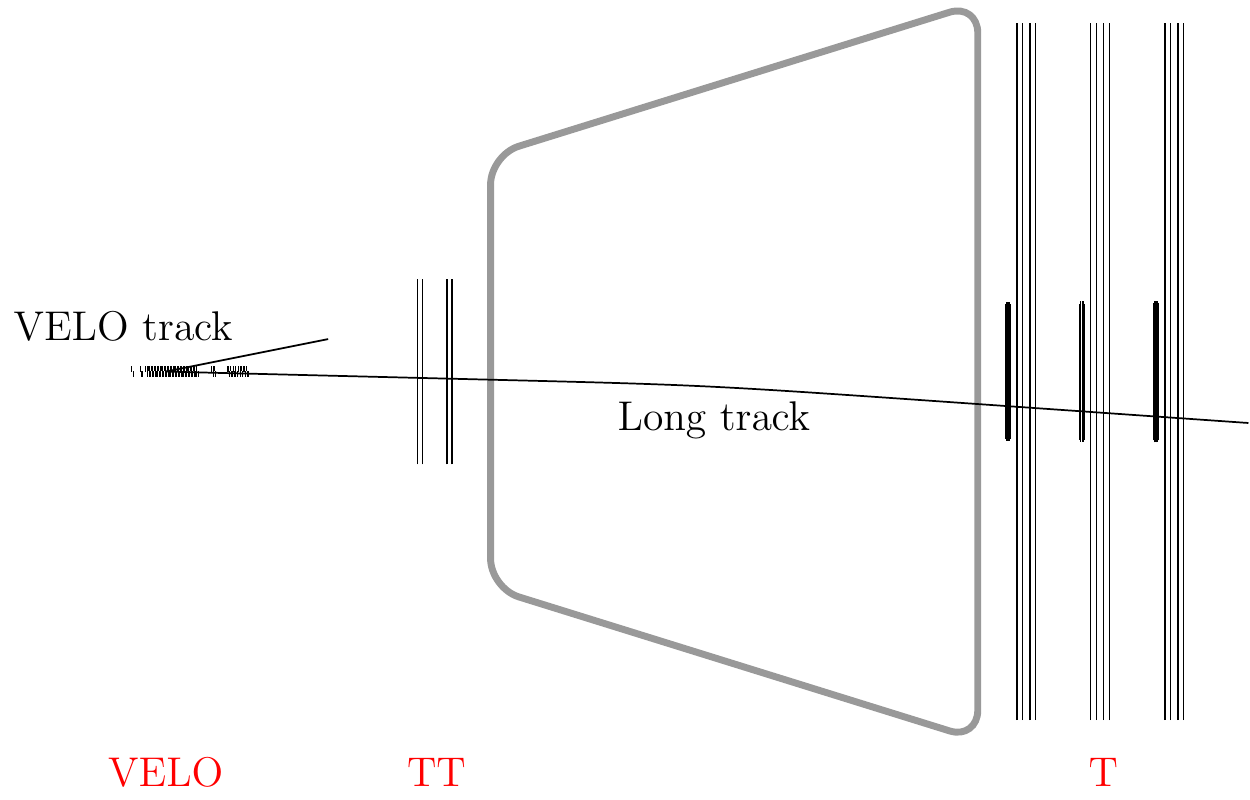}
	\caption{Schematic view of the different track types relevant for this paper, along with a schematic of the essential tracking detectors of the LHCb detector, showing the VELO detector, TT detector, magnet and the downstream tracking stations. The names of the tracking detectors are indicated in red below the figure. }
	\label{fig:experiment_track_types}
\end{figure}

Figure~\ref{fig:experiment_track_types} shows an overview of the relevant track types used in this paper: \velo tracks, composed of at least three hits in the \velo detector; and long tracks, which have hits in the VELO detector, T stations and optionally the TT detector. The momentum is inferred from the track's curvature in the magnetic field. Most physics analyses make use of long tracks, as they provide the best momentum and angular resolution among all of the track types. The \velo resides outside the magnetic field, and charged particles follow an approximately straight trajectory through the detector. Therefore, a \velo track only provides angular information about the particle's initial trajectory, and no information about the particle's absolute momentum. The reconstruction of long tracks starts with the reconstruction of \velo tracks, which are subsequently promoted by dedicated algorithms to long tracks~\cite{LHCb-2007-015,LHCb-2007-020}. Consequently, all long tracks must also satisfy the conditions of the \velo track reconstruction. 

The trigger~\cite{LHCb-DP-2012-004} consists of a
hardware stage, based on information from the calorimeter and muon
systems, followed by a software stage which performs a full event reconstruction. This analysis uses specific trigger selections deployed for 2017 data taking which build the tag-and-probe candidates required for the analysis at the trigger level. In order to minimize potential correlations between the tag and probe sides of the signal candidate, the trigger selection is based on a minimal set of criteria required to reduce the output rate to an acceptable level.

Simulated events are used to validate the proposed method. 
  In the simulation, $pp$ collisions are generated using
\pythia~\cite{Sjostrand:2007gs} 
 with a specific \lhcb
configuration~\cite{LHCb-PROC-2010-056}.  Decays of unstable particles
are described by \evtgen~\cite{Lange:2001uf}, in which final-state
radiation is generated using \photos~\cite{Golonka:2005pn}. The
interaction of the generated particles with the detector, and its response,
are implemented using the \geant
toolkit~\cite{Allison:2006ve, *Agostinelli:2002hh} as described in
Ref.~\cite{LHCb-PROC-2011-006}. 
\section{Method}
\label{sect:method}
\begin{figure}[]
\includegraphics[width=0.49\textwidth]{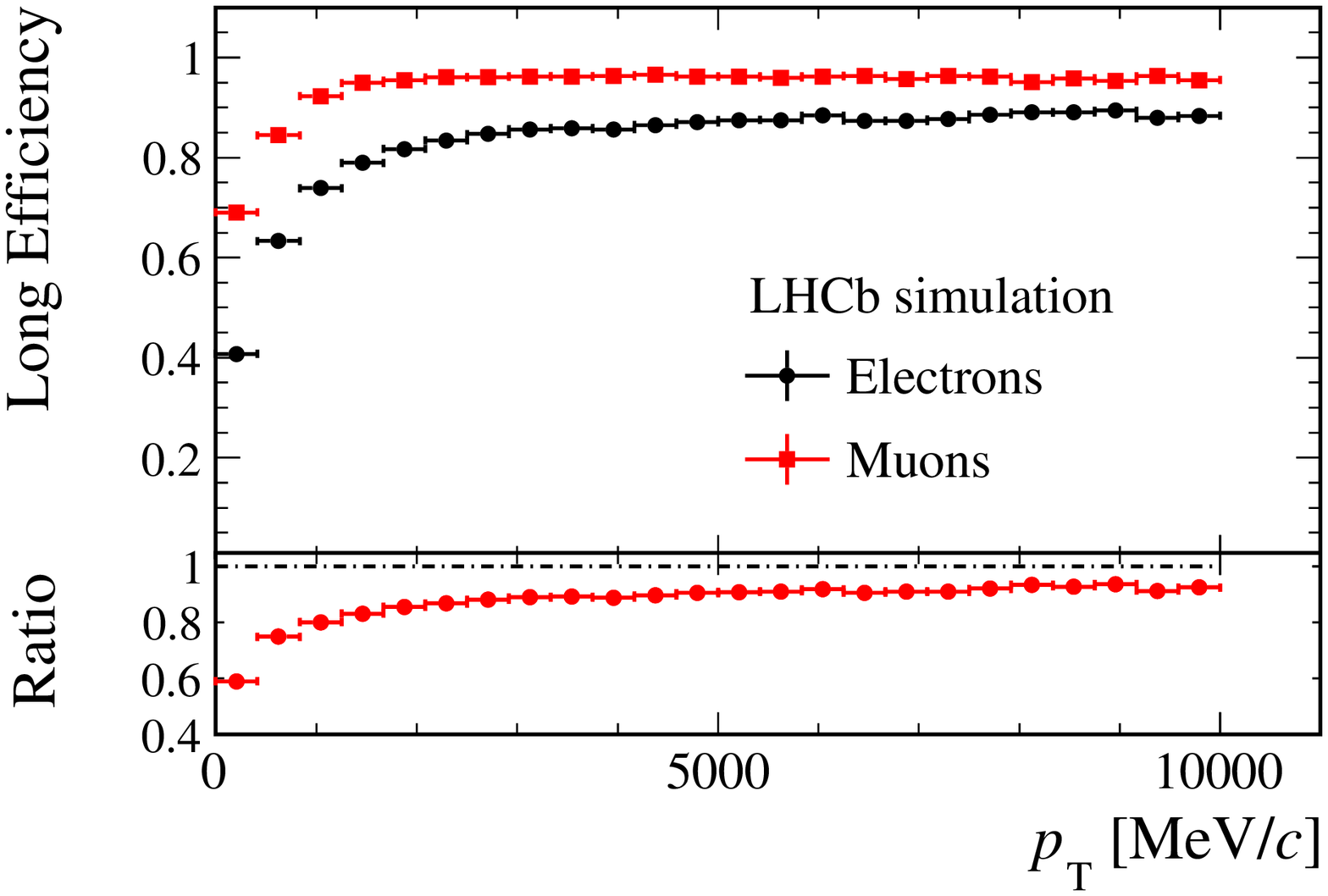}
\includegraphics[width=0.49\textwidth]{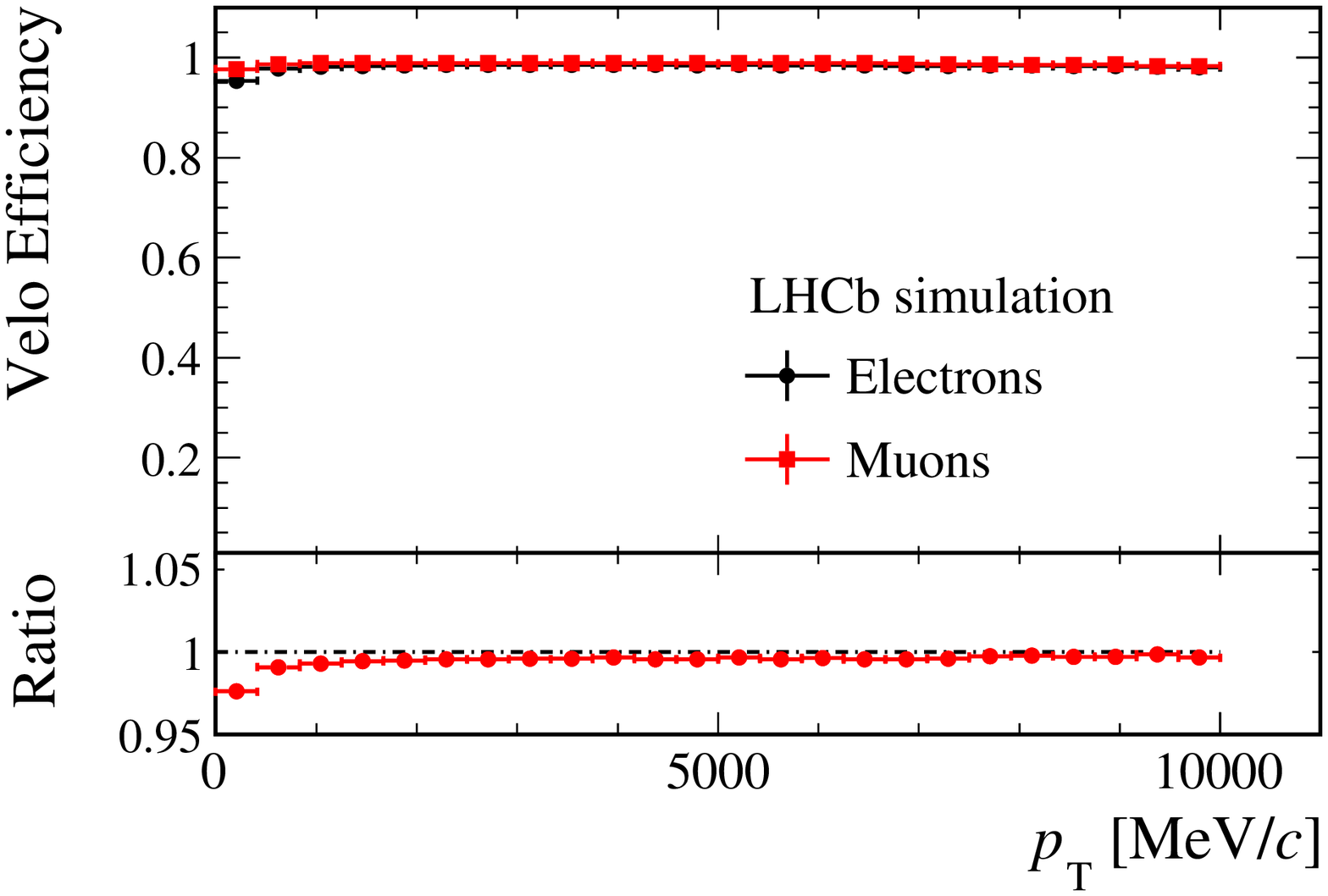}
\caption{The simulated LHCb (left) long-track and (right) VELO-track reconstruction efficiency for electrons and muons produced in  $B^{+}\to J/\psi(\to \ellell) K^+$  decays as a function of the lepton's transverse momentum.}
\label{fig:electronrecoeff}
\end{figure}

The long-track reconstruction efficiency for electrons is significantly different with respect to muons, as shown in Fig.~\ref{fig:electronrecoeff}. While the \velo-track reconstruction efficiencies are similar and close to $100\%$, electrons that undergo radiative energy losses before the dipole magnet can be deflected outside of the downstream tracker acceptance. This acceptance inefficiency is significantly larger than the inefficiency of the pattern-recognition algorithms. For a measurement of the long-track reconstruction efficiency for electrons it is therefore required to rely only on subdetectors upstream of the dipole magnet.

This paper describes the efficiency of reconstructing electrons as long tracks, provided that they have been reconstructed as \velo tracks. Therefore, the presented method does not account for the efficiency of the \velo-track reconstruction. A calibration procedure exists already for the \velo-track reconstruction~\cite{LHCb-DP-2013-002}, which also takes into account the ageing of the detector. In addition, the \velo detector material is described well in the simulation, with a relative precision on $x/X_{0}$ of $6\%$~\cite{LHCb-DP-2014-001}. Combining the  calibration for muons with the detector simulation, the \velo-track reconstruction efficiency for electrons can be determined with a systematic uncertainty of approximately $0.06\%$ due to the simulated material density, which is small in comparison to the current statistical uncertainty.

This tag-and-probe method uses $\Bp\to\jpsi(\to\ep\en)\Kp$ decays to determine the efficiency of reconstructing electrons as long tracks. The decay products cover the momentum spectrum of interest to most analyses studying properties of heavy flavour in LHCb, with a typical average transverse momentum of a few $\gevc$. One of the electrons and the kaon from the $\Bp\to\jpsi(\to\ep\en)\Kp$ decay are reconstructed as long tracks and define the tag candidate, while the other electron is reconstructed in the \velo and defines the probe candidate. The angular information of the \velo track combined with the information of the tag particles is sufficient to distinguish genuine signal decays from backgrounds, as illustrated in Sect.~\ref{sect:momentum_and_invariant_mass_resolution}. The long-track reconstruction efficiency for electrons is obtained by matching the \velo track to a long track. The probe track is defined as matched if a long track is found which has at least $70\%$ of the clusters in the \velo detector in common, with a minimum of $6$ shared clusters on a track.\footnote{Both the \velo-track reconstruction and the long-track reconstruction require a minimum of 6 \velo clusters on a track.} As an additional quality requirement, the reconstructed long track is required to have the correct electric charge associated, \ie opposite to the charge of the tag electron. 

\subsection{Efficiency parametrisation}
The efficiency of the long-track reconstruction depends on the absolute momentum of the electron and the amount of traversed material. The electron reconstruction efficiency is therefore measured in bins of pseudorapidity, $\eta$, and transverse momentum, $\pt$. The bin sizes are chosen to minimise the variation of the efficiency within the kinematic bin, based on the simulated behaviour of the electron reconstruction efficiency, with a lower limit on the bin size due to the limited sample size. 

Within the VELO, a striking variation in the amount of traversed material is present due to the RF Foil, an aluminium foil which separates the detector and beam vacua. Particles which travel parallel to the RF-foil can traverse an additional $10\%$ of radiation length in comparison to other particles, and thus are treated separately. This region is defined as two regions in the azimuthal angle, $\phi$, of the probe electron: $|\phi - \pi/2| < \pi/8$ and $|\phi + \pi/2| < \pi/8$.

\subsection{Momentum and invariant-mass resolution}
\label{sect:momentum_and_invariant_mass_resolution}
The momentum of the probe electron is inferred from kinematic and geometric constraints in the signal production and decay chain.
This  momentum estimate is used in the determination of the $\ep\en\Kp$ invariant mass, whose invariant-mass distribution is used to separate signal from backgrounds. In addition, an accurate estimate of the probe's momentum is required to parametrise the reconstruction efficiency. Different kinematic constraints are applied in these two cases. 

\begin{figure}[t]
\centering
\includegraphics[width=0.49\textwidth]{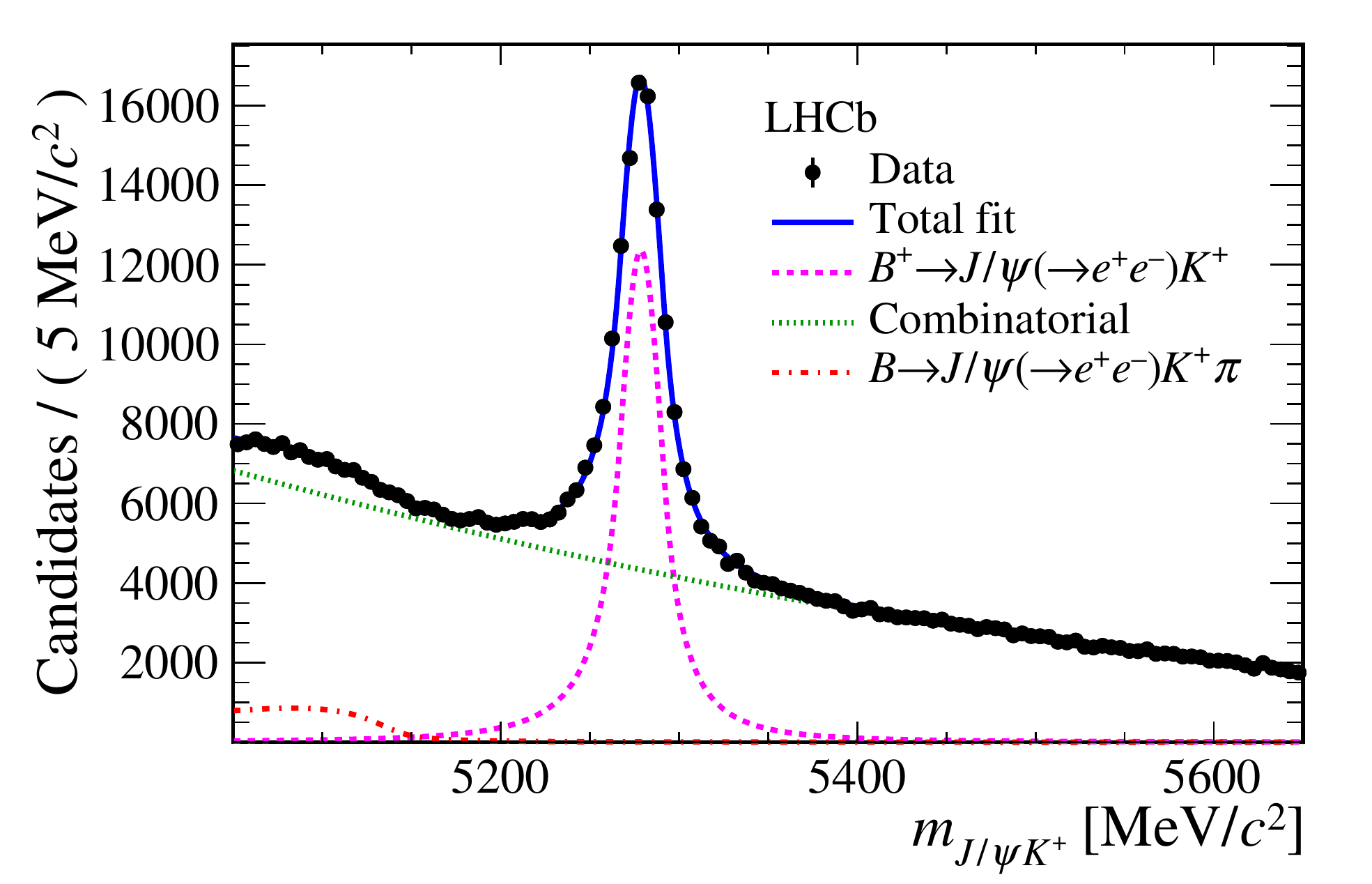}
\includegraphics[width=0.49\textwidth]{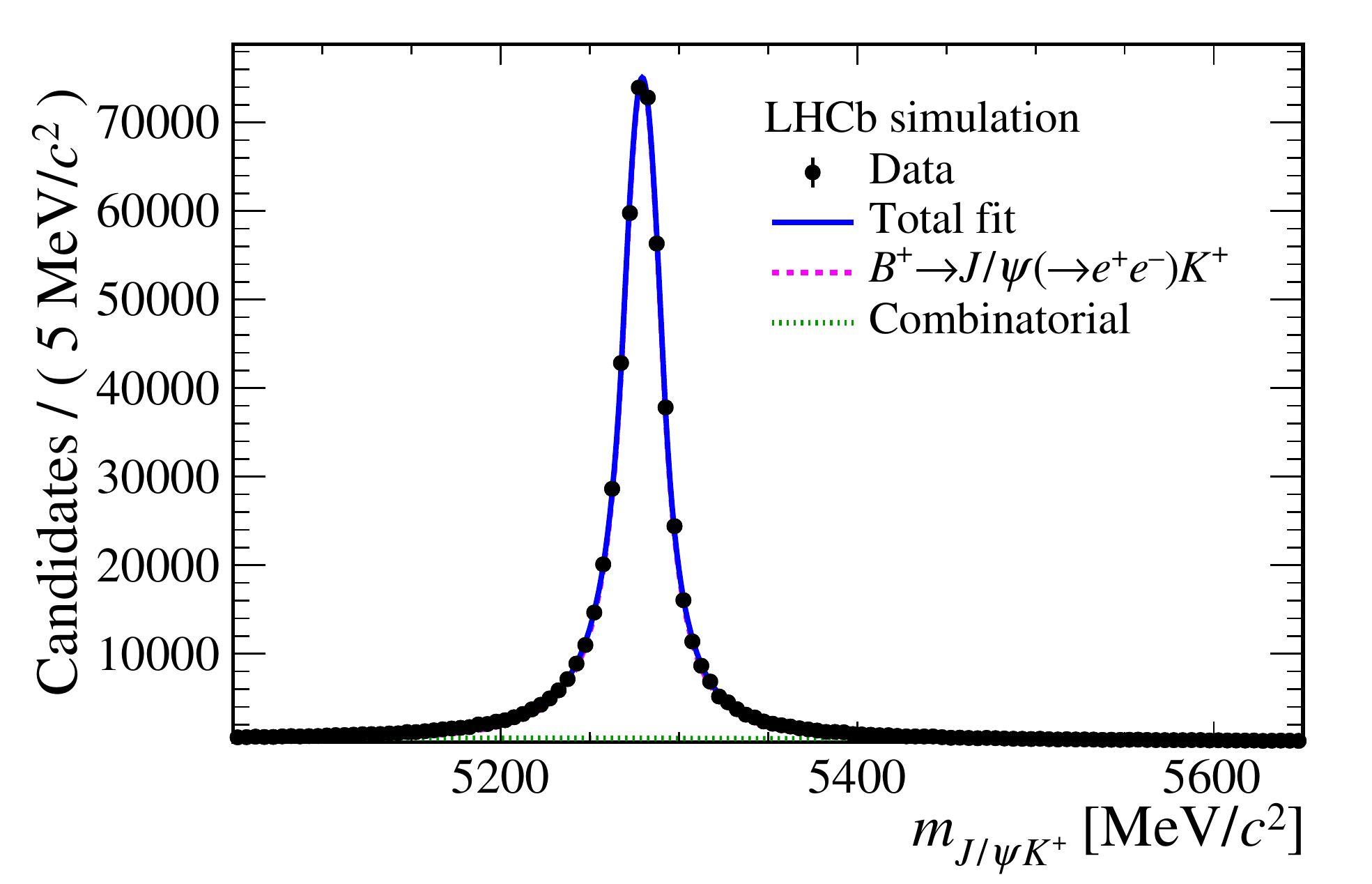}
\caption{Distribution of the constrained invariant mass for (left) data and (right) simulated events, together with an example of a fit to this distribution. The simulated events contain at least one signal decay, resulting in a higher signal purity than is observed in data.}\label{fig:fit_shape_example}
\end{figure}

To separate genuine signal candidates from background, a fit is performed to the $\ep\en\Kp$ invariant-mass distribution. The $\ep\en\Kp$ invariant mass is computed by constraining the invariant mass of the $\en\ep$ combination to the known value for the $\jpsi$ mass~\cite{PDG2018}. An example of the distribution of the resulting invariant mass, $m_{\jpsi \Kp}$, is shown in Fig.~\ref{fig:fit_shape_example}. The shape of the signal peak is reproduced well in the simulation, shown in the right panel of Fig.~\ref{fig:fit_shape_example}. The signal peak is fitted with a Gaussian function with power-law tails on both sides. The tail parameters are extracted from simulation, and fixed in a fit to the data. The random-track background is described empirically using a second-degree polynomial, whose parameters are left to float in a fit to the data. In addition to random combinations of unrelated tracks, also partially reconstructed decays of the type $\Bp\to\jpsi(\to\ep\en)\Kstar$ contribute in the invariant-mass spectrum. This contribution is described using an Argus function~\cite{ARGUS} convoluted with a resolution function, whose parameters are left to float in a fit to the data. All parameters are determined separately for each bin of the probe electron kinematics.

The momentum estimation of the probe track for the efficiency parametrisation follows a similar, but more involved, procedure. The tag electron also suffers from energy loss as it traverses the detector, which worsens the momentum resolution of the tag electron and, consequently, affects the probe momentum inferred from the $\jpsi$ mass constraint. To further improve the momentum resolution of the probe electron, a kinematic fit is performed to the decay tree, in which both the probe and tag electron's absolute momentum are determined by constraining the $\ep\en\Kp$ invariant mass to the known $\Bp$ mass, and simultaneously constraining the $\ep\en$ invariant mass to the known $\jpsi$ mass. The resulting relative momentum resolution, as evaluated using simulation, is approximately $7\%$, with little dependence on the momentum of the probe electron. This momentum estimate is used when defining the kinematic binning for the efficiency measurement. Migration effects due to the momentum resolution are treated as a systematic uncertainty and addressed in Sect.~\ref{sect:systematic_errors}.

\section{Data set and selection} 
This analysis uses  $\sqrt{s}=13\tev$ $pp$ collision data recorded by the \lhcb detector in 2017, corresponding to an integrated luminosity of $1.3\invfb$. The majority of the event selection for this analysis is implemented in a dedicated trigger line. This line combines a track, which is likely an electron, with another track, which is likely to be a kaon, to form together with the probe a vertex which is displaced by at least $4\mm$ from the nearest PV. Both tag tracks must form a vertex with a significant displacement with respect to any PV. The tag electron must satisfy $\pt>2.5\gevc$, while the tag kaon must satisfy $\pt\geq500\mevc$. This tag combination is then combined with a \velo track (the probe candidate) which is also required to have a significant displacement with respect to any PV. To reduce the large background of random-track combinations, a selection is made based on an approximation of the invariant mass of the $\ep\en\Kp$ combination. For this, the momentum of the probe electron, $p_{\mathrm{probe}}$ is approximated as
\begin{equation}
	\label{eq:approximated_mass_constraint}
	p_{\mathrm{probe}} = \frac{1}{2} \frac{m_{\jpsi}^{2} - 2 m_{\mathrm{e}}^{2}}{E_{\mathrm{e,tag}} - p_{\mathrm{e,tag}} \cos\theta},
\end{equation}where $m_{\jpsi}$ denotes the known mass of a $\jpsi$ meson~\cite{PDG2018}, $m_{e}$ denotes the mass of an electron, $E_{\mathrm{e,tag}}$ ($p_{\mathrm{e,tag}}$) denotes the energy (momentum) of the tag electron, and $\theta$ denotes the opening angle between the two electrons in the vertex detector. The PV that fits best to the resulting flight direction of the \Bp candidate is used henceforth as the associated PV.

To avoid biases in the measured efficiency, all considered events must have passed the trigger independently of the long-track reconstruction for the probe electron. After tracks have been reconstructed, trigger decisions are associated with reconstructed tracks or energy deposits~\cite{LHCb-PUB-2014-039}. This allows one to ensure that the event triggers accepted the event due to the tag particles. As a consequence of these requirements, the tag electron must have a significant transverse momentum, which is reflected in the $\pt\geq2.5\gevc$ requirement implemented in the selection described earlier.

An offline selection is used to further increase the purity of candidates passing the trigger selection. The angle between the $\Bp$ momentum vector, calculated using the probe momentum estimation from the $\jpsi$ mass constraint, and the reconstructed flight direction is required to be at most $5.5\mrad$. In addition, the probe electron is required to have an impact parameter with respect to any reconstructed PV of at least $0.2\mm$.

Simulated events are processed in the same way. About $10^{5}$ $\Bp$ decays are selected in data, and $5.7 \times 10^5$ in simulated events.

\section{Efficiency extraction}
\label{sect:efficiency_fit}
\begin{figure}[]
\centering
\includegraphics[width=0.48\textwidth]{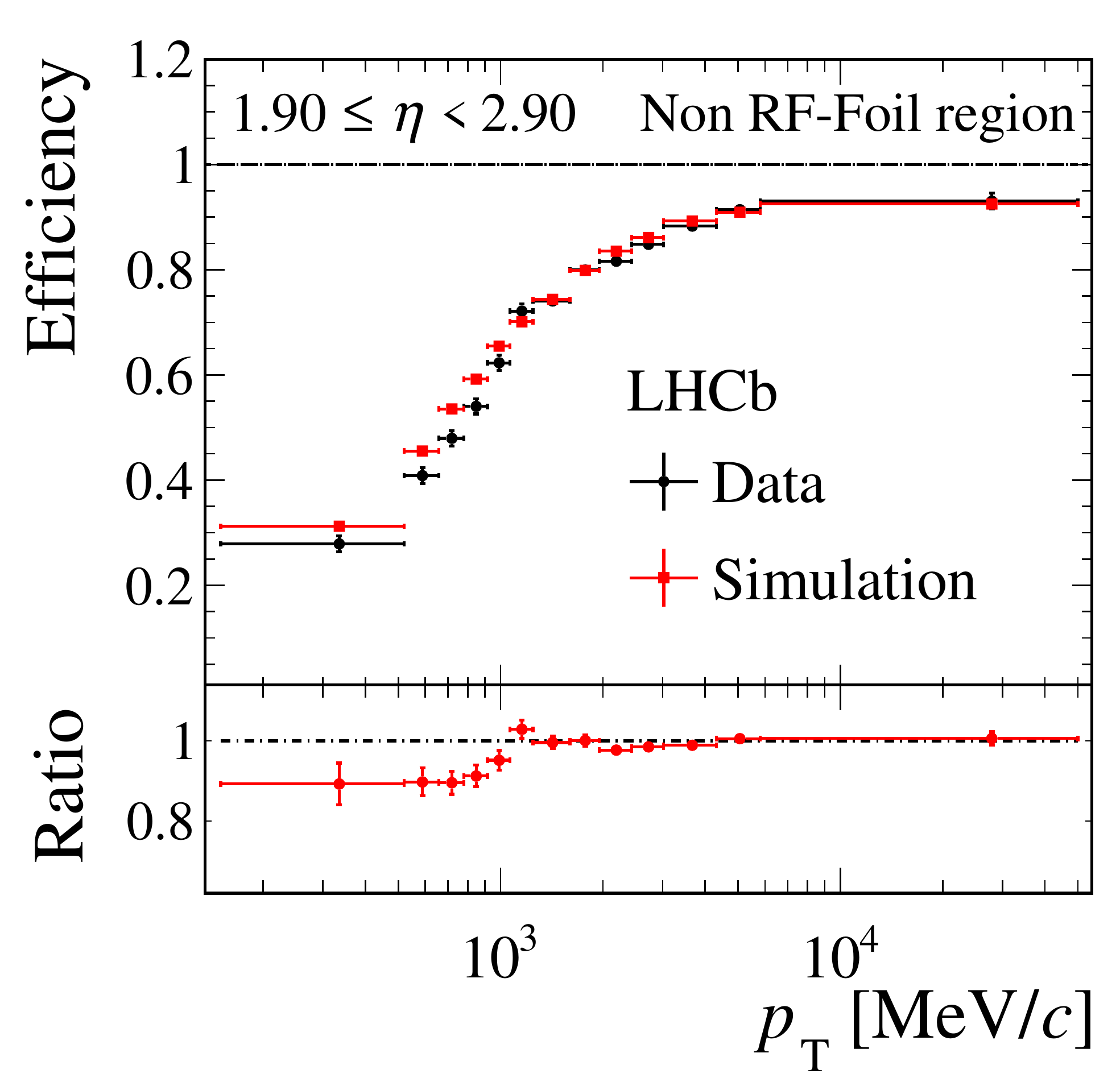}
\includegraphics[width=0.48\textwidth]{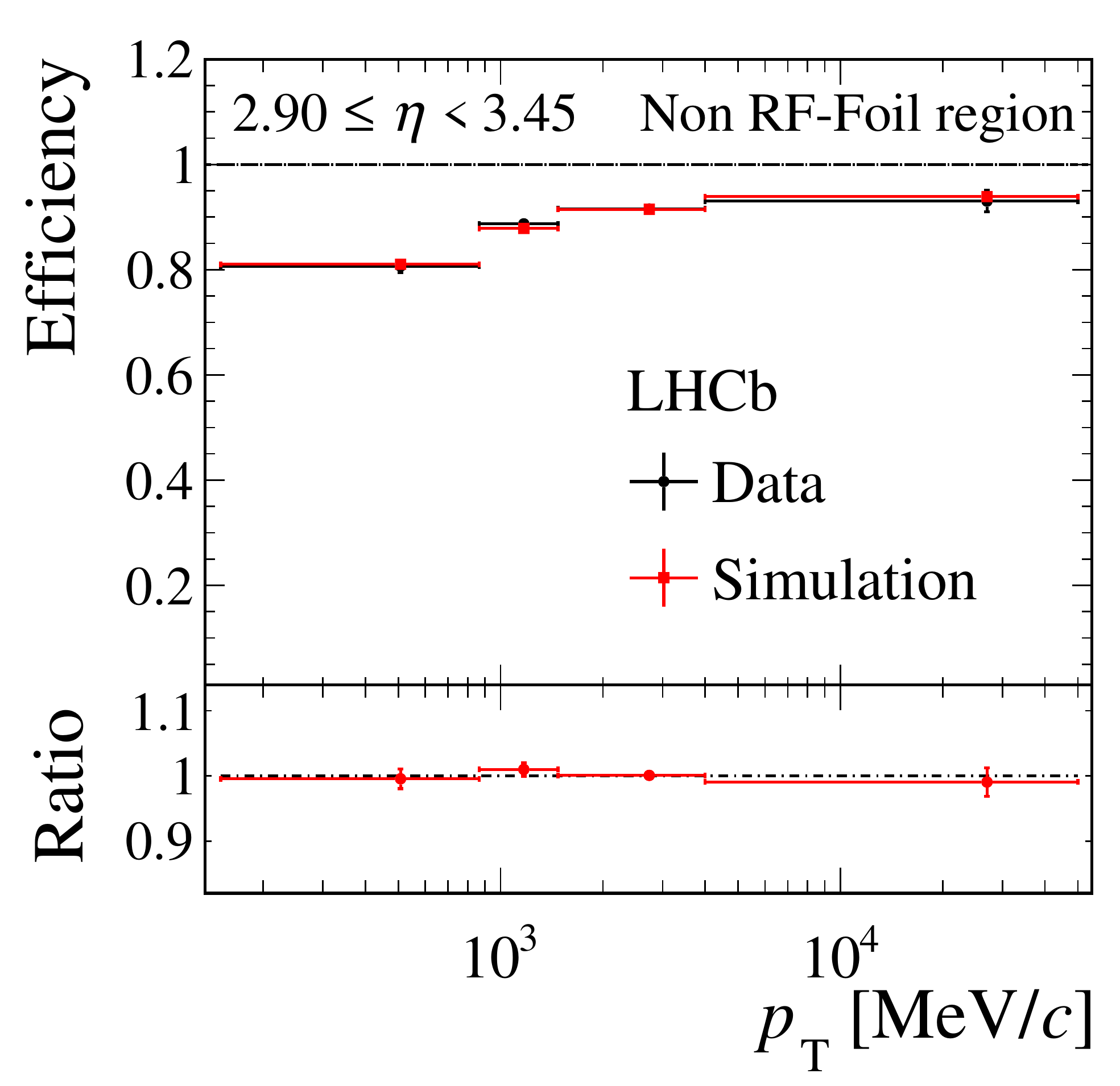}\\
\includegraphics[width=0.48\textwidth]{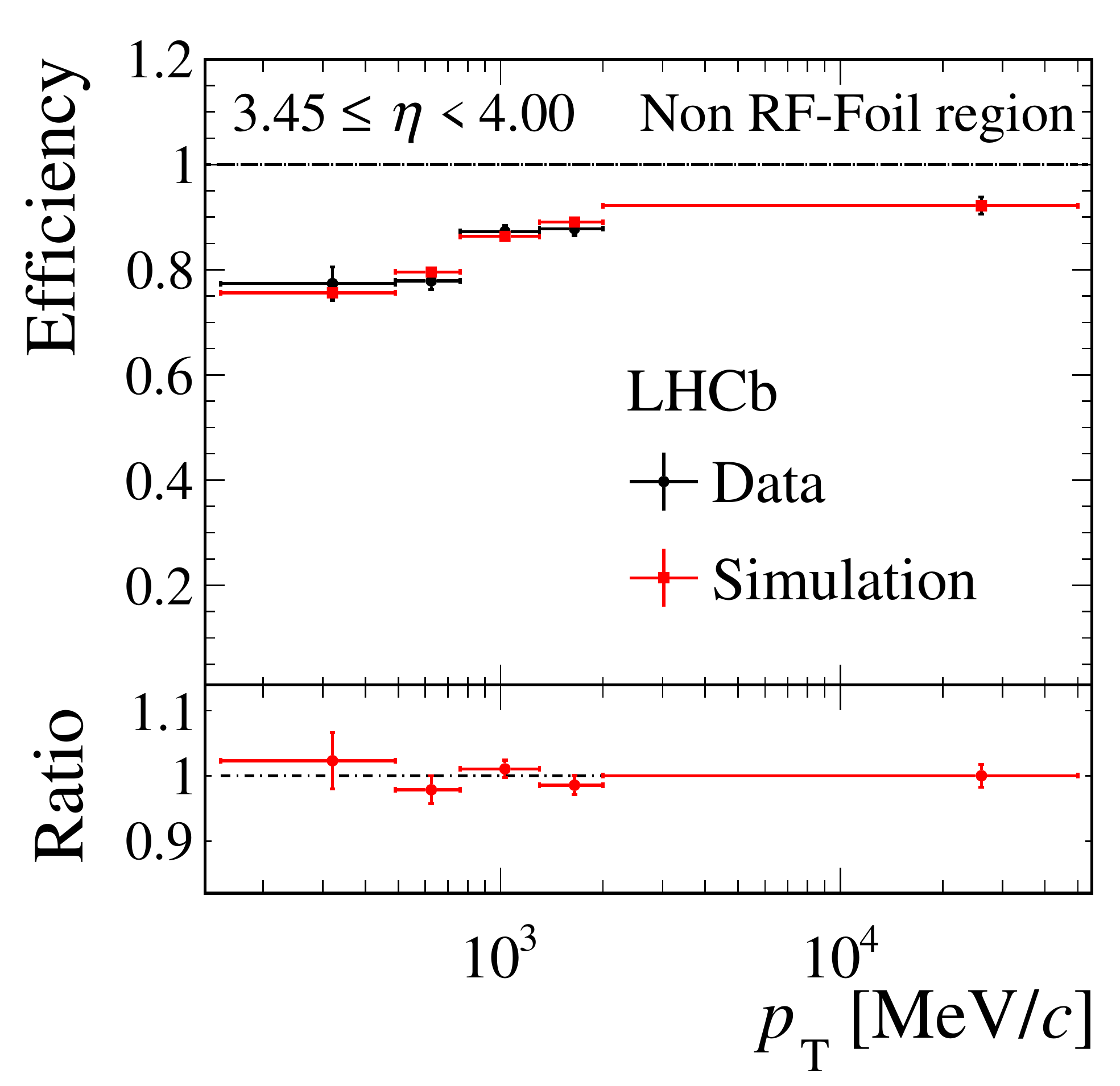}
\includegraphics[width=0.48\textwidth]{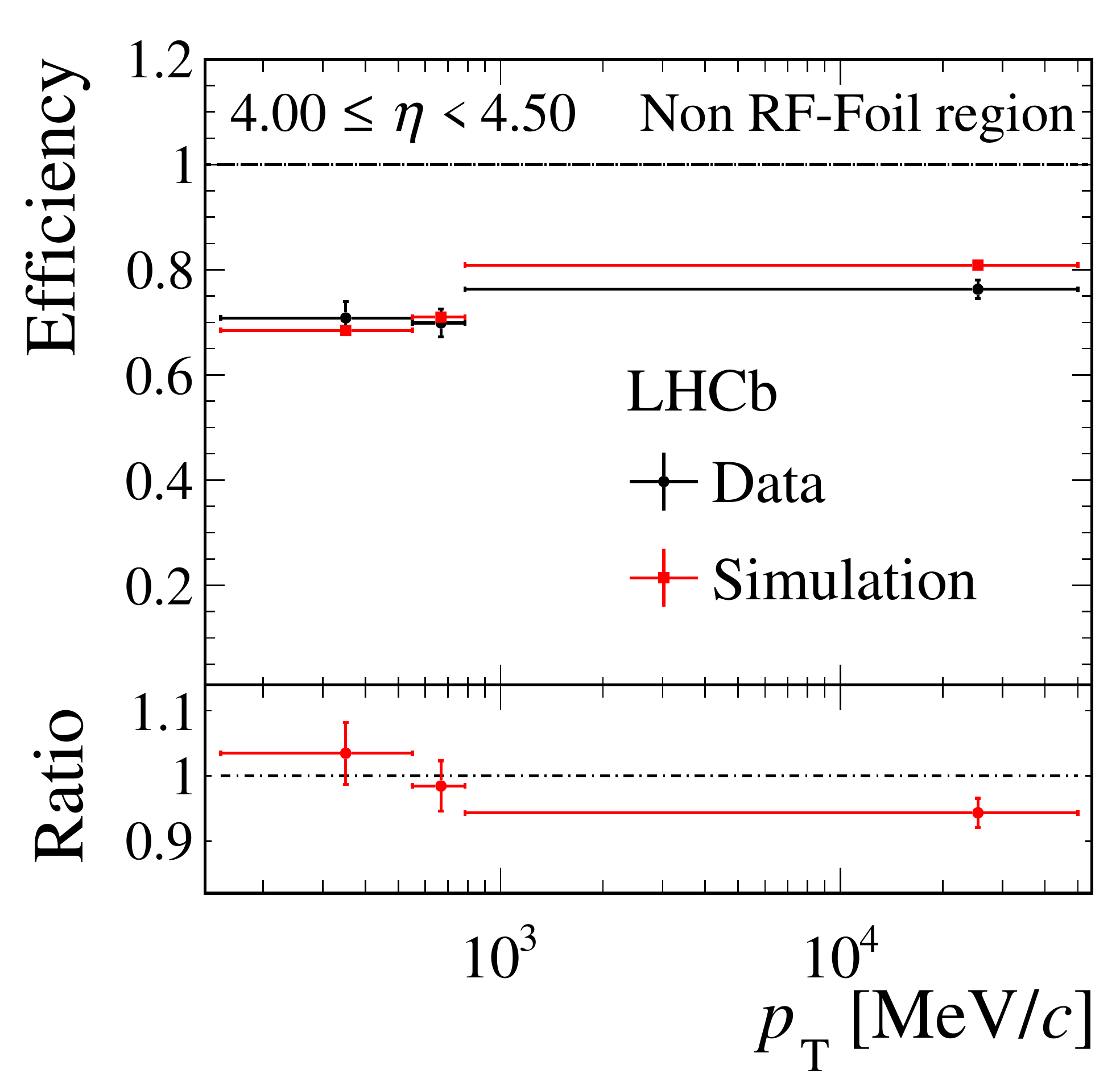}
\caption{Measured electron detection efficiency in each $\pt$, $\eta$ bin in data and simulation, for electrons which do not travel parallel to the RF-foil. The ratios between the results from data and simulation are shown for each bin below. The error bars on the ratio include both the statistical and systematic uncertainties, while those for the absolute efficiencies represent only statistical uncertainties. }
\label{fig:dataeffele_phi0}
\end{figure}

\begin{figure}[]
\centering
\includegraphics[width=0.48\textwidth]{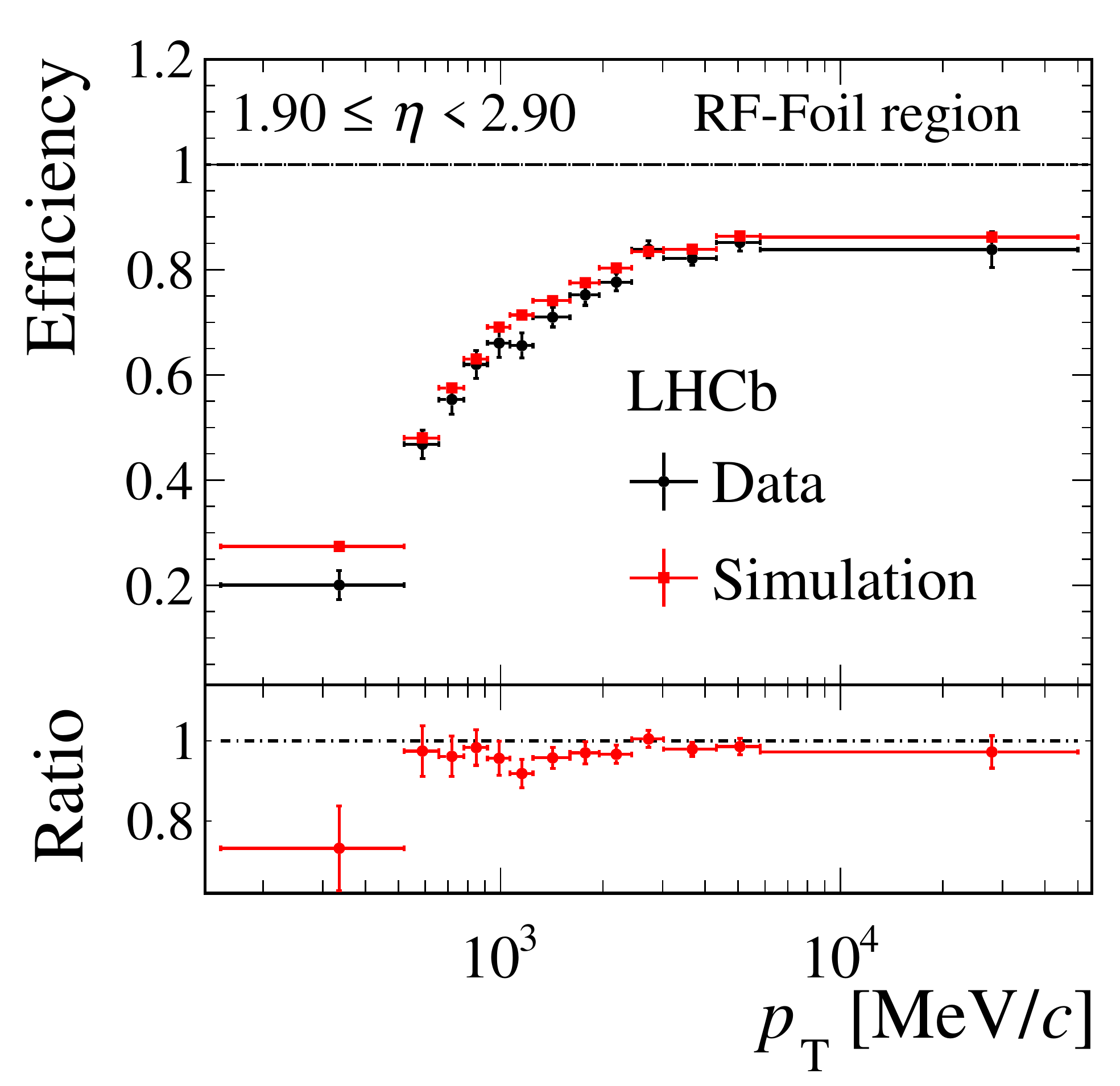}
\includegraphics[width=0.48\textwidth]{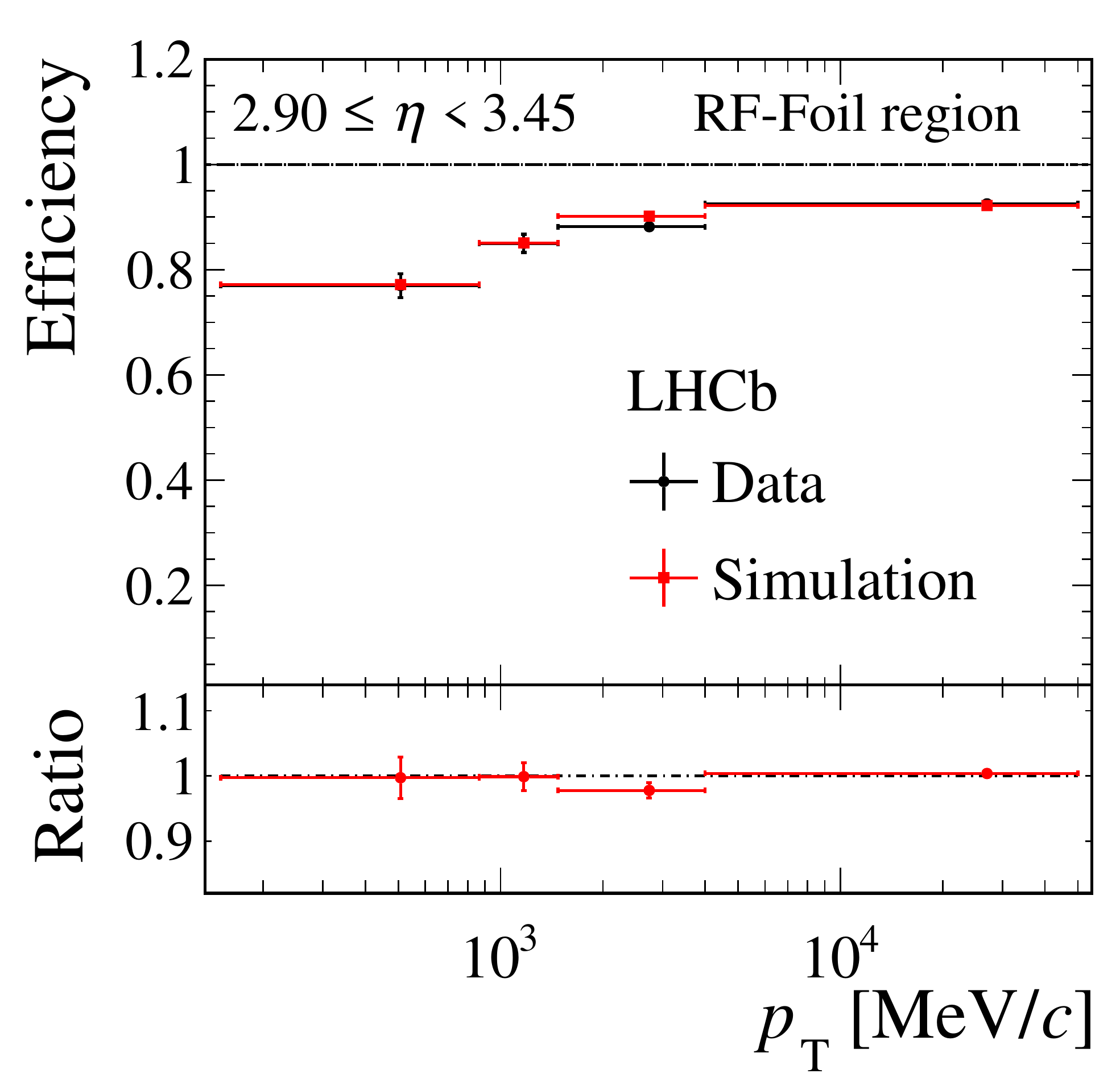}\\
\includegraphics[width=0.48\textwidth]{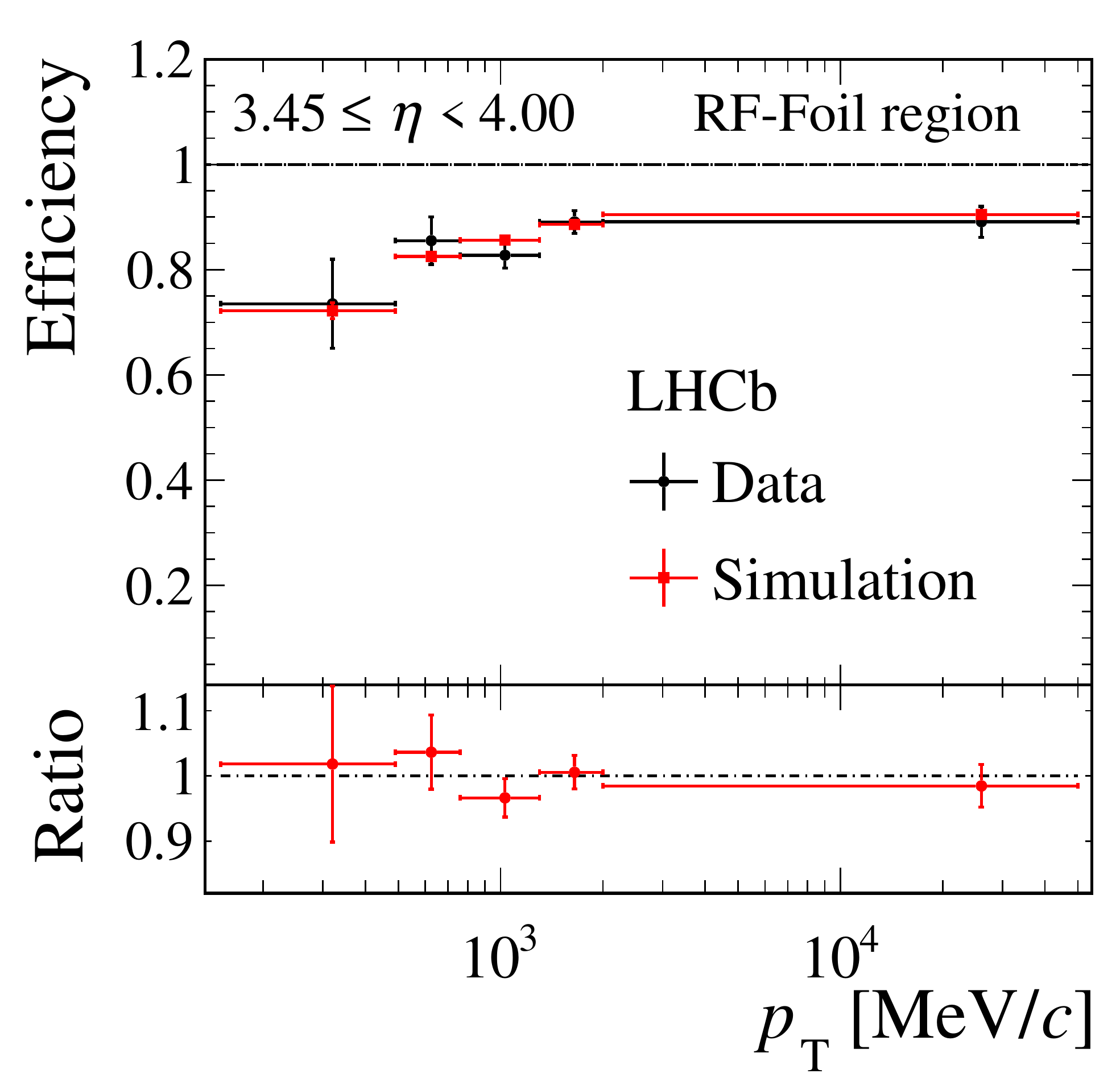}
\includegraphics[width=0.48\textwidth]{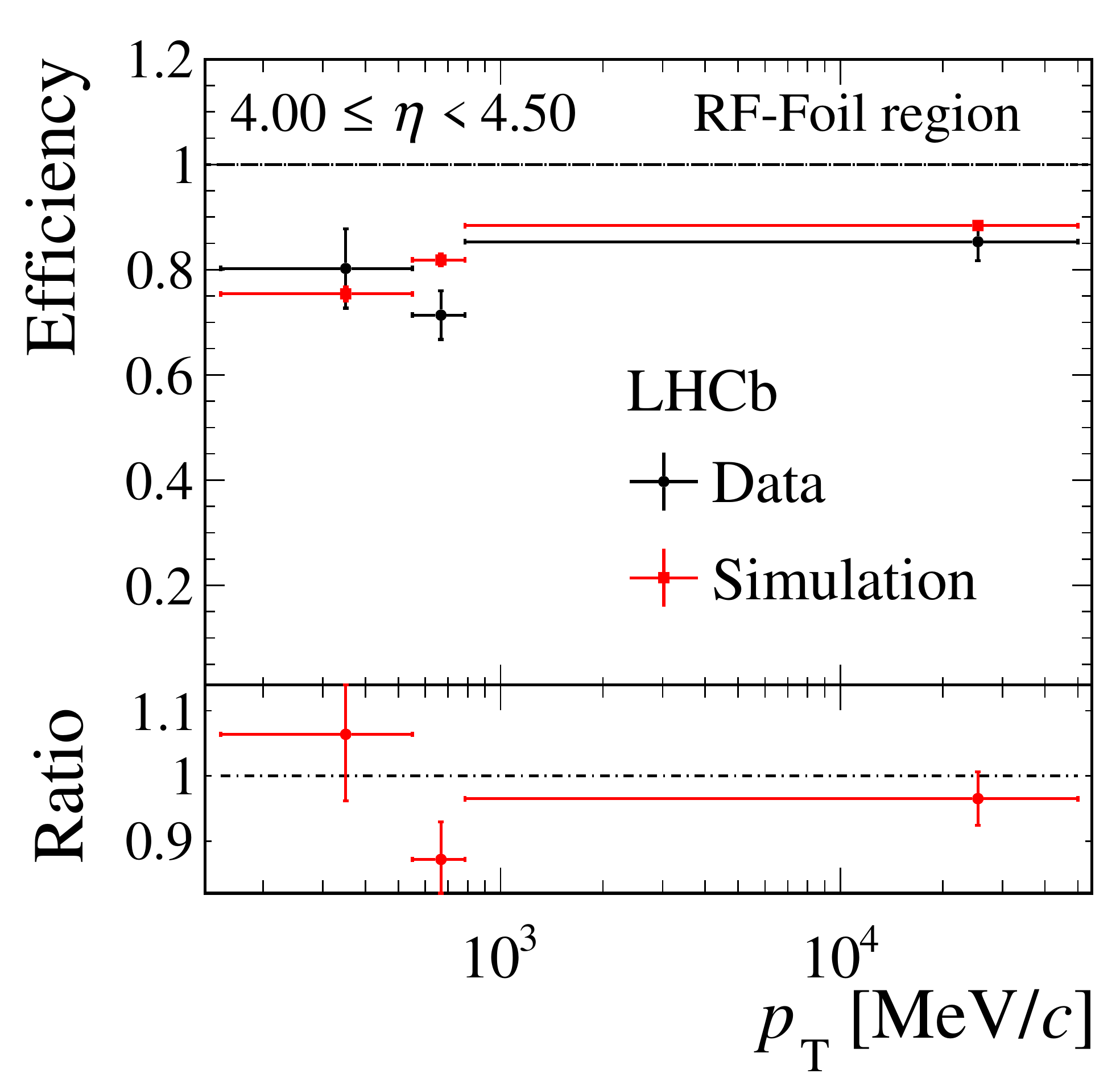}
	\clearpage
\caption{Measured electron detection efficiency in each $\pt$, $\eta$ bin in data and simulation, for electrons which travel parallel to the RF-foil. The ratios between the results from data and simulation are shown for each bin below. The error bars on the ratio include both the statistical and systematic uncertainties, while those for the absolute efficiencies represent only statistical uncertainties.}
\label{fig:dataeffele_phi1}
\end{figure}

The reconstruction efficiency of electrons is determined as the fraction of reconstructed decays where the probe track is  matched to a long track with the expected electric charge. The selected candidates are divided into pass and fail categories based on this matching criterion. The efficiency is subsequently measured using a simultaneous one-dimensional, unbinned maximum-likelihood fit to the inferred mass of the $\Bp$ meson candidate in both categories. To minimise differences in the treatment of the pass and fail categories, the $\Bp$ meson mass is calculated using the inferred probe momentum in both cases. 
The resulting efficiencies in simulation and data along with their statistical uncertainties are shown in Figs.~\ref{fig:dataeffele_phi0} and~\ref{fig:dataeffele_phi1}. 

There is a good agreement between simulation and data in most kinematic bins, with an average deviation below $5\%$. At low $\pt$ and $\eta$ the efficiency in data is consistently lower than what is observed in simulation. This is attributed to a difference in detection efficiency for electrons with a low momentum, $p \leq 3 \gevc$. The selection requirements made on the tag candidate, and in particular the tag electron \pt requirement, limit our ability to study the probe efficiency in this region in more detail. Most \lhcb physics analyses~\cite{LHCb-DP-2019-001,LHCb-DP-2013-001} require a minimal reconstructed momentum of $3 \gevc$, such that the practical impact of this efficiency difference is mitigated.

\subsection{Fake track contribution}
Not all of the long tracks correspond to genuine particle trajectories. Long tracks which are composed of the probe particle's \velo segment and an unrelated segment in the downstream tracking stations appear as long tracks which have the reconstructed angle of the searched-for probe particle, but an incorrect estimate of the absolute momentum. The reconstruction of such fake tracks happens more frequently for electrons than for other particle species, as the conversion of bremsstrahlung photons introduces additional charged particles in the downstream tracking stations which appear compatible with the \velo segment. 

In physics analyses, which often restrict themselves to long tracks, the momentum resolution is already worsened due to the energy loss of the probe electron. This energy loss impacts the resolution of invariant masses, such that they no longer provide a clear distinction between genuine and fake tracks, unlike for other particle species. To provide a general performance number, contributions from fake tracks are included in the efficiency presented in this paper. However, this definition can be changed when evaluating the efficiency for a specific physics analysis in order to reflect the applied selection criteria.

\subsection{Efficiency ratio and validation}
The truth-level hit information present in simulated events allows for another definition of the track reconstruction efficiency. Using this information, a charged particle is considered reconstructed when at least 70\% of the hits on its track are associated to the same true simulated particle. Using simulated events, this hit-based efficiency agrees within $2\%$ with the efficiency measured with the tag-and-probe method for all kinematic bins, after the fake-track contribution is subtracted. To mitigate the influence of this discrepancy, the ratio between the efficiencies in data and simulation is considered,
\[
	\mathrm {Ratio} = \frac{\varepsilon_\mathrm{data}}{\varepsilon_\mathrm{simulation}}.
\]The efficiency ratios act as correction factors to simulated events and, as the ratio is most robust to systematic uncertainties, are considered the main results of this method. The values for the efficiency ratios are shown in Figs.~\ref{fig:dataeffele_phi0},~\ref{fig:dataeffele_phi1}.\nowidow 

\section{Stability}
\label{sect:stability}
\begin{figure}[b]
    \begin{center}
    \includegraphics[width=0.48\textwidth]{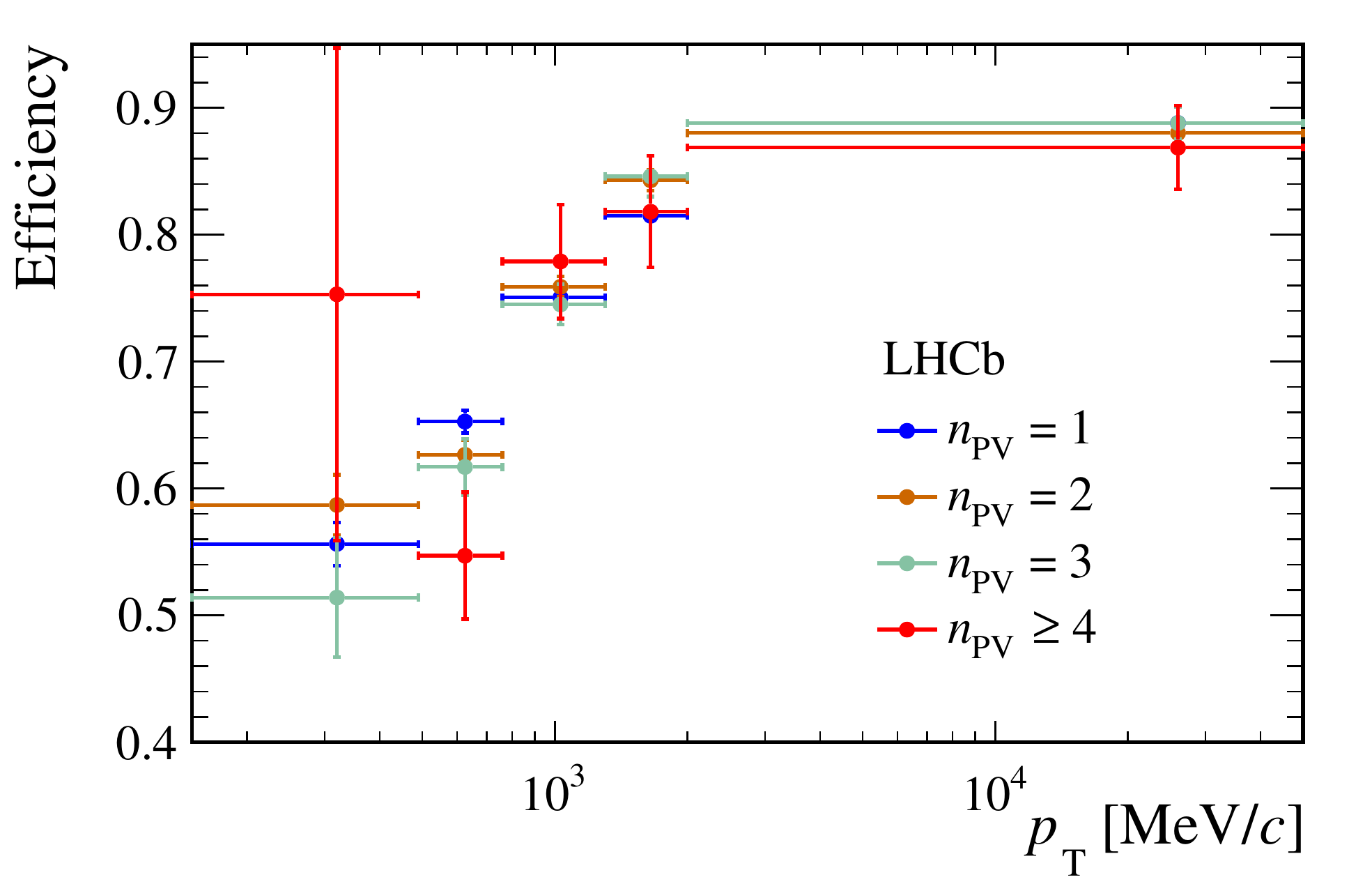}
    \includegraphics[width=0.48\textwidth]{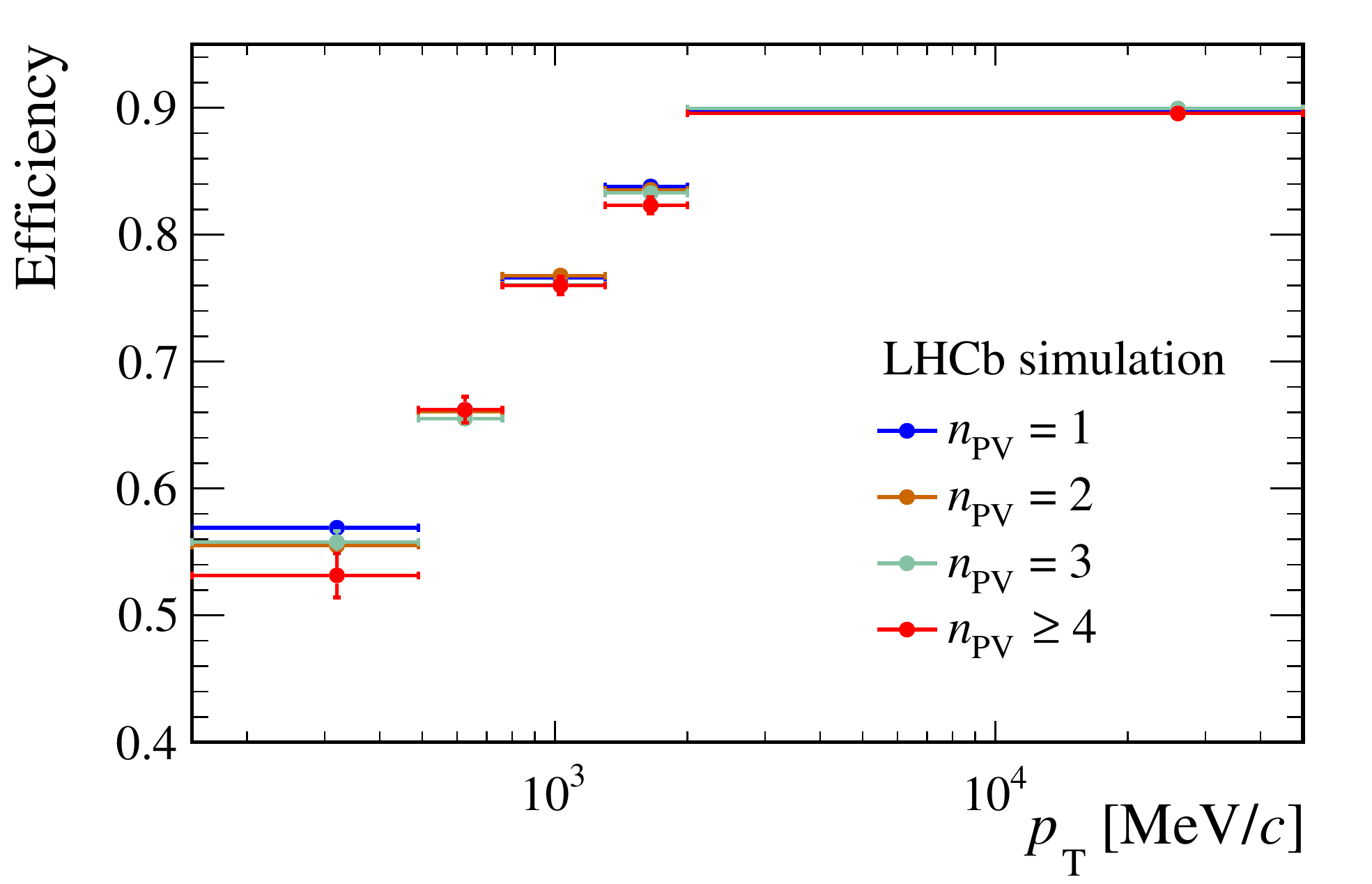}
        \caption{Efficiency in bins of number of primary vertices and the transverse momentum of the probe electron for (left) data and (right) simulation. The uncertainties shown are statistical only. }
        \label{fig:syst:systematics_npv}
         \end{center}
\end{figure}

The reconstruction efficiency for muons as long tracks is known to depend on the detector occupancy~\cite{LHCb-DP-2013-002}. In simulation, the reconstruction efficiency of electrons also shows a weak dependence on the number of additional $pp$ interactions in the same bunch crossing (pile-up), measured as the number of reconstructed PVs in the event. To validate that this dependence is of similar order in data, the electron detection efficiency is measured in bins of number of reconstructed PVs multiplicity. The results are shown in Fig.~\ref{fig:syst:systematics_npv}. As expected, no dependencies on the number of PVs are observed with the current statistical precision.

\section{Systematic uncertainties}
\label{sect:systematic_errors}
The impact of common systematic uncertainties between simulation and data is mitigated by the consideration of the efficiency ratio. In particular, imperfections in the detector simulation affecting the method can lead to a residual systematic uncertainty. This section describes all of the evaluated systematic uncertainties for the efficiency ratio. 

To apply this calibration in physics analyses it is assumed that, with the correct parametrisation, the detection efficiency is independent of the decay topology. To validate this assumption, selection criteria are varied around their nominal values. For those selection criteria which are well modelled, the impact is estimated from the deviations observed in simulated events. These variations impact the absolute efficiency by $0.39\%$ on average, and are included in the systematic uncertainties for the ratio. The influence of selection criteria involving quantities related to the track reconstruction on the inferred efficiency is studied in both data and simulated events, verifying the description in the simulation. While the absolute efficiencies depend on the applied selection criteria, the efficiency ratio remains unaffected. 

The signal model in the invariant-mass distribution is shared between pass and fail categories, in addition to data and simulation. Therefore, the ratio is expected to be largely insensitive to the details of the fit to the invariant-mass distributions. When an alternative model~\cite{Santos:2013gra} is used to describe the signal by determining the model per pass and fail categories, a variation in the ratio is observed of at most $0.5\%$, which confirms this hypothesis. This contribution is included in the total systematic uncertainties. 

The limited momentum resolution can affect the inferred difference in the efficiency curves and the ratio between results in simulation and data. As the resulting uncertainty is smaller than the statistical uncertainties, the bin migration is treated as a systematic error, and no effort is taken to unfold the $\pt$ distribution. The resulting systematic uncertainty is $\order(0.15\%)$ on average, but varies up to $2.5\%$ in a region of phase-space with large statistical uncertainties on the measured ratios and a strong variation in the efficiency.

In the efficiency ratio a dependency on the implementation of the $\Bp$ production model in simulation is introduced. This dependency is largely mitigated by the binning applied, but in-bin variations can still be present. A comparison of the probe kinematics between data and simulation showed no significant differences thus far when considered in each bin, provided the limited statistics available, and no systematic uncertainty is assigned for this effect. With more data, a weighting procedure can be applied to remove such differences.

\begin{figure}[] 
    \begin{center}
    \includegraphics[width=0.9\textwidth]{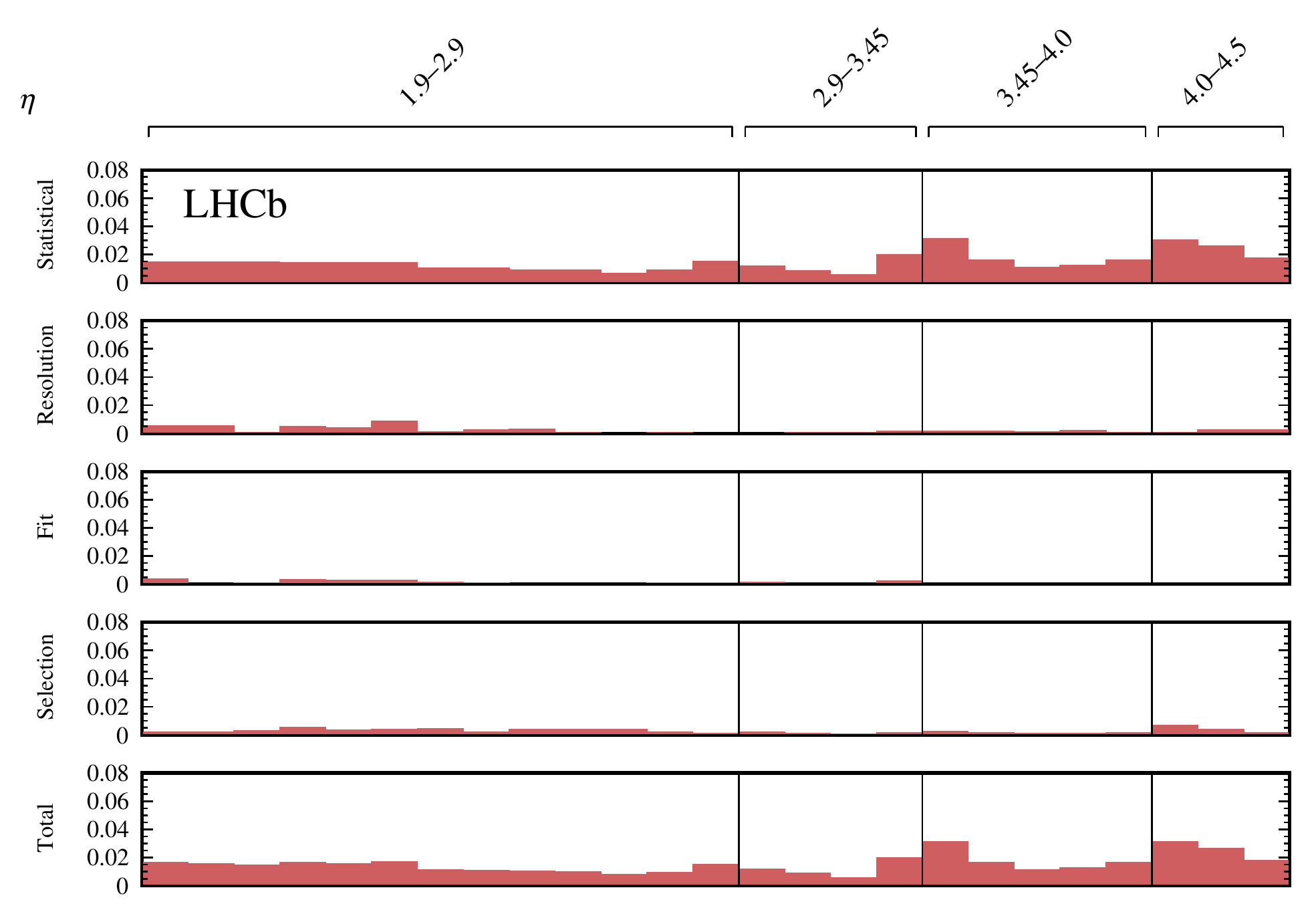}
	\\
    \includegraphics[width=0.9\textwidth]{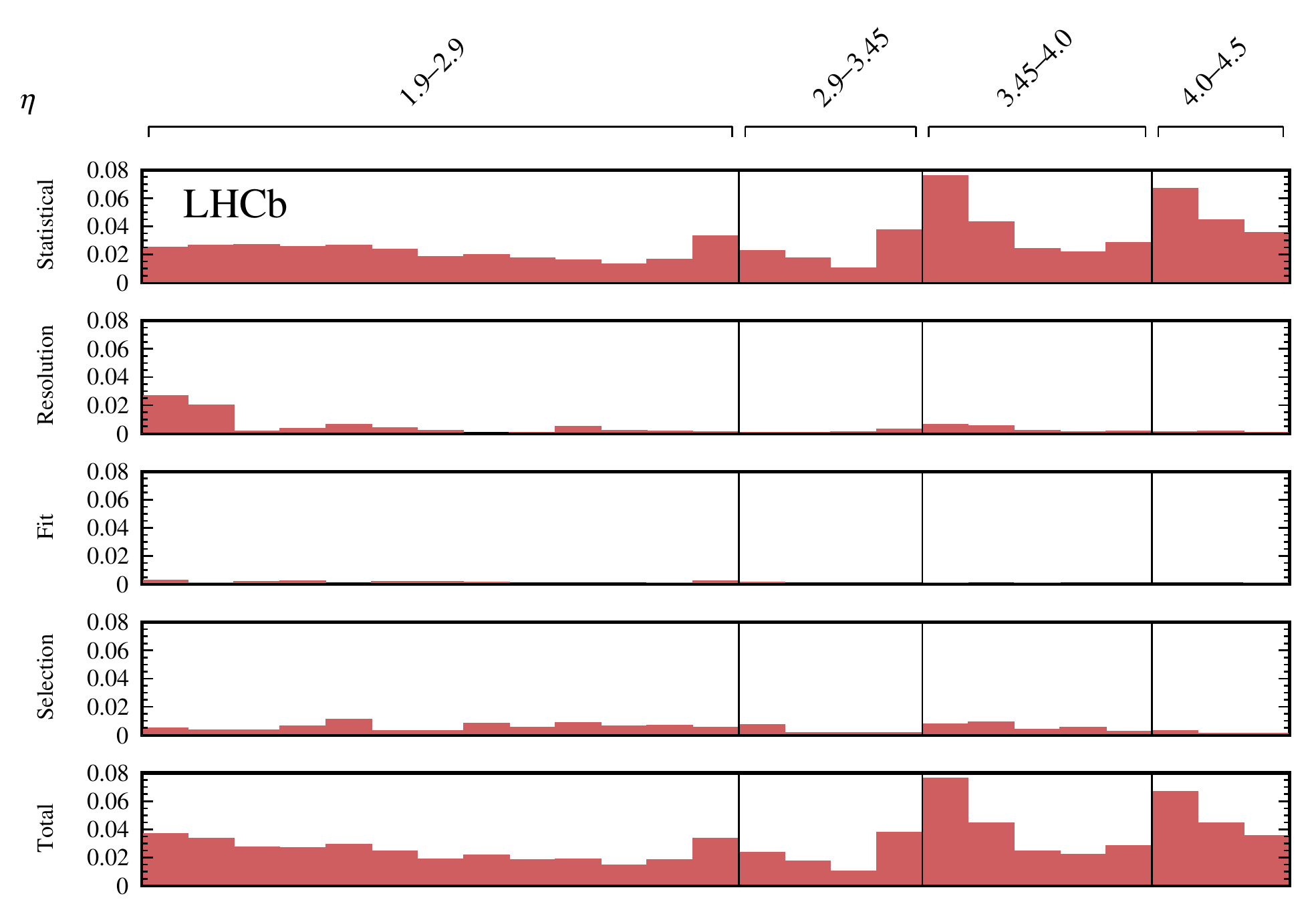}
    \caption{Overview of all uncertainties on the efficiency ratio per bin of \pt and $\eta$ of the probe electron for (bottom) candidates which traverse parallel to the RF-foil and (top) the rest of the phase-space. The $x$-axis corresponds to the \pt bin number within the $\eta$ range. }
    \label{fig:syst:totalsystematics}
    \end{center}
\end{figure}

All uncertainties on the efficiency ratios, including the different studied systematic uncertainties and the  statistical uncertainties, are shown in Fig.~\ref{fig:syst:totalsystematics} for each bin in kinematics.

\section{Conclusion}
\label{sect:conclusion}
A novel technique to determine detection and reconstruction efficiencies for electrons at \lhcb has been presented. Results using part of the data recorded in 2017 show a large variation of this efficiency over phase-space, comparable but not equal to the results from simulation. To provide a suitable correction for physics analyses, only the ratio between the reconstruction efficiency of electrons in data and simulation is relevant, in which  systematic uncertainties are mitigated. The average systematic uncertainty on this ratio is $0.6\%$ per track and varies with phase-space.

\section*{Acknowledgements}
%
%
\noindent We express our gratitude to our colleagues in the CERN
accelerator departments for the excellent performance of the LHC. We
thank the technical and administrative staff at the LHCb
institutes.
We acknowledge support from CERN and from the national agencies:
CAPES, CNPq, FAPERJ and FINEP (Brazil); 
MOST and NSFC (China); 
CNRS/IN2P3 (France); 
BMBF, DFG and MPG (Germany); 
INFN (Italy); 
NWO (Netherlands); 
MNiSW and NCN (Poland); 
MEN/IFA (Romania); 
MSHE (Russia); 
MinECo (Spain); 
SNSF and SER (Switzerland); 
NASU (Ukraine); 
STFC (United Kingdom); 
DOE NP and NSF (USA).
We acknowledge the computing resources that are provided by CERN, IN2P3
(France), KIT and DESY (Germany), INFN (Italy), SURF (Netherlands),
PIC (Spain), GridPP (United Kingdom), RRCKI and Yandex
LLC (Russia), CSCS (Switzerland), IFIN-HH (Romania), CBPF (Brazil),
PL-GRID (Poland) and OSC (USA).
We are indebted to the communities behind the multiple open-source
software packages on which we depend.
Individual groups or members have received support from
AvH Foundation (Germany);
EPLANET, Marie Sk\l{}odowska-Curie Actions and ERC (European Union);
ANR, Labex P2IO and OCEVU, and R\'{e}gion Auvergne-Rh\^{o}ne-Alpes (France);
Key Research Program of Frontier Sciences of CAS, CAS PIFI, and the Thousand Talents Program (China);
RFBR, RSF and Yandex LLC (Russia);
GVA, XuntaGal and GENCAT (Spain);
the Royal Society
and the Leverhulme Trust (United Kingdom).

\addcontentsline{toc}{section}{References}
\bibliographystyle{LHCb}
\bibliography{main,standard,LHCb-PAPER,LHCb-CONF,LHCb-DP,LHCb-TDR}

\newpage
\centerline
{\large\bf LHCb collaboration}
\begin
{flushleft}
\small
R.~Aaij$^{30}$,
C.~Abell{\'a}n~Beteta$^{47}$,
T.~Ackernley$^{57}$,
B.~Adeva$^{44}$,
M.~Adinolfi$^{51}$,
H.~Afsharnia$^{8}$,
C.A.~Aidala$^{78}$,
S.~Aiola$^{24}$,
Z.~Ajaltouni$^{8}$,
S.~Akar$^{62}$
% ...

{\footnotesize \it

% ...
$ ^{8}$Universit{\'e} Clermont Auvergne, CNRS/IN2P3, LPC, Clermont-Ferrand, France\\
% ...
$ ^{24}$INFN Sezione di Milano, Milano, Italy\\
% ...
$ ^{30}$Nikhef National Institute for Subatomic Physics, Amsterdam, Netherlands\\
% ...
$ ^{44}$Instituto Galego de F{\'\i}sica de Altas Enerx{\'\i}as (IGFAE), Universidade de Santiago de Compostela, Santiago de Compostela, Spain\\
% ...
$ ^{47}$Physik-Institut, Universit{\"a}t Z{\"u}rich, Z{\"u}rich, Switzerland\\
% ...
$ ^{51}$H.H. Wills Physics Laboratory, University of Bristol, Bristol, United Kingdom\\
% ...
$ ^{57}$Oliver Lodge Laboratory, University of Liverpool, Liverpool, United Kingdom\\
% ...
$ ^{62}$University of Cincinnati, Cincinnati, OH, United States\\
% ...
$ ^{65}$Syracuse University, Syracuse, NY, United States\\
% ...
$ ^{78}$University of Michigan, Ann Arbor, United States, associated to $^{65}$\\
\bigskip
% ...
}
\end{flushleft}

\end{document}